\documentclass[emulateapj,natbib]{emulateapj}

\begin{document}
\title{Deep Thermal Infrared Imaging of HR 8799 \lowercase{bcde}: New Atmospheric Constraints and Limits on a Fifth Planet}

\author{Thayne Currie\altaffilmark{1,11}, Adam Burrows\altaffilmark{2}, 
Julien H. Girard\altaffilmark{3},
Ryan Cloutier\altaffilmark{1},
Misato Fukagawa\altaffilmark{4}, 
Satoko Sorahana\altaffilmark{5},
Marc Kuchner\altaffilmark{6}, 
Scott J. Kenyon\altaffilmark{7},
Nikku Madhusudhan\altaffilmark{8}, 
Yoichi Itoh\altaffilmark{9},
Ray Jayawardhana\altaffilmark{1},
Soko Matsumura\altaffilmark{10},
Tae-Soo Pyo\altaffilmark{11}
}
\altaffiltext{1}{Department of Astronomy and Astrophysics,University of Toronto}
\altaffiltext{2}{Department of Astrophysical Sciences, Princeton University}
\altaffiltext{3}{European Southern Observatory}

\altaffiltext{4}{Osaka University}
\altaffiltext{5}{University of Tokyo}
\altaffiltext{6}{NASA-Goddard Space Flight Center}
\altaffiltext{7}{Harvard-Smithsonian Center for Astrophysics}
\altaffiltext{8}{Institute for Astronomy, University of Cambridge}
\altaffiltext{9}{University of Hyago}
\altaffiltext{10}{University of Dundee}
\altaffiltext{11}{National Astronomical Observatory of Japan}
\begin{abstract}
We present new $L^\prime$ (3.8 $\mu m$) and Br-$\alpha$ (4.05 $\mu m$) data and reprocessed archival $L^\prime$ data for the young, planet-hosting star 
HR 8799 obtained with Keck/NIRC2, VLT/NaCo and Subaru/IRCS.
We detect all four HR 8799 planets in each dataset at a moderate to high 
signal-to-noise (SNR $\gtrsim$ 6-15).
 We fail to identify a fifth planet, ``HR 8799 f", at $r$ $<$ 15 $AU$ at a 5-$\sigma$ confidence level: one suggestive, marginally significant residual at 0\farcs{}2 is most likely a PSF artifact.
 Assuming companion ages of 30 $Myr$ and the Baraffe (Spiegel \& Burrows) 
planet cooling models, we rule out an HR 8799 f with mass of 
5 $M_{J}$ (7 $M_{J}$), 7 $M_{J}$ (10 $M_{J}$), and 12 $M_{J}$ (13 $M_{J}$) at $r_{proj}$ $\sim$ 12 $AU$, 9 $AU$, and 5 $AU$, respectively.
 All four HR 8799 planets have red early T dwarf-like $L^\prime$ - [4.05] colors, suggesting that their 
SEDs peak in between the $L^\prime$ and $M^\prime$ broadband filters.  
We find no statistically significant difference in HR 8799 cde's colors.
Atmosphere models assuming thick, patchy clouds appear to better match HR 8799 bcde's photometry than models assuming a uniform cloud layer.
While non-equilibrium carbon chemistry is required to explain HR 8799 bc's photometry/spectra, evidence for it from HR 8799 de's photometry is weaker.   
Future, deep IR spectroscopy/spectrophotometry with the Gemini Planet Imager, SCExAO/CHARIS, and other facilities may clarify whether 
the planets are chemically similar or heterogeneous.

\end{abstract}
\section{Introduction}
The four directly-imaged planets around the young A-type star HR 8799 \citep[HR 8799 bcde;][]{Marois2008,Marois2010a} provide a crucial reference point for understanding the 
 physical properties of gas giants soon after their formation.      
HR 8799 bcde have best-estimated masses of $\sim$ 5--7 $M_{J}$ \citep{Marois2010a,Currie2011a,Sudol2012} and are located at projected separations of $r_{proj}$ $\approx$ 15--70 $AU$, in between a warm, inner dust belt ($r_{inner}$ $\approx$ 6-12 $AU$) and an outer, Kuiper belt-like structure at $r_{outer}$ $\approx$ 90 $AU$ \citep{Su2009}.   Due to HR 8799's higher luminosity, the planets receive about as much energy as the gas/ice giant planets in our solar system receive from the Sun. Thus, the HR 8799 planetary system may resemble a young, scaled-up version of our own solar system \citep{Marois2010a}.

 HR 8799 bcde highlight key, ubiquitous features of young gas giants several times the mass of Jupiter.  Compared to the locus of field brown dwarfs, the planets appear ``red" or ``under luminous" at the shortest infrared wavelengths \citep[i.e. 1--1.6 $\mu m$][]{Marois2008,Currie2011a}.  This trend is due to the planets being cloudier/dustier than field substellar objects at the same effective temperatures \citep[i.e. $T_{\rm{eff}}$ $\sim$ 800--1200 $K$][]{Currie2011a}.  The clouds may not be uniformly distributed but instead ``patchy" \citep[][]{Currie2011a} with 10-50\% of the visible surface covered by thinner clouds \citep[see also][]{Skemer2012,Skemer2014}.   At longer wavelengths, some planets show evidence for non-equilibrium carbon chemistry, exhibiting weak to negligible methane absorption at 3.3 $\mu m$  and enhanced $CO$ absorption at 5 $\mu m$ \citep{Hinz2010,Galicher2011,Skemer2012}.    Near-infrared spectra for HR 8799 bc reveal molecular species in the planets' atmospheres and additional evidence for non-equilibrium carbon chemistry \citep{Bowler2010,Oppenheimer2013,Konopacky2013}.  
 
 Both thick clouds and non-equilibrium carbon chemistry likely have the same physical origin: the planets' low surface gravities.  Such low gravities \citep[log(g) $\sim$ 4 instead of $\sim$ 5][]{Currie2011a,Madhusudhan2011,Konopacky2013} yield temperature-pressure profiles more characteristic of hotter, cloudier dwarfs and enhance vertical mixing in the planets' atmospheres \citep[e.g.][]{Madhusudhan2011, Marley2012}.  Other young, directly-imaged planets/planet-mass companions (e.g. HD 95086 b, $\beta$ Pic b, ROXs 42Bb, and 2M 1207 B) also show evidence for these trends \citep{Galicher2014, Currie2013,Currie2014a,Barman2011,Skemer2011}.  

However, HR 8799 bcde may yet differ in important ways.  HR 8799 b has a lower luminosity and, by inference, a lower mass than the inner three planets.  Best-fit atmosphere models adopting physically realistic radii (i.e. 1--1.4 $R_{J}$)  and yielding estimated masses consistent with dynamical stability limits ($M_{bcde}$ $\le$ 7 $M_{J}$) imply that HR 8799 b's photosphere is $\sim$ 100-200 $K$ cooler than HR 8799 cde \citep[][]{Currie2011a,Madhusudhan2011}.     

Recent spectroscopy of multiple HR 8799 planets from Project 1640 imply that HR 8799 cde may have different chemistries \citep{Oppenheimer2013}.    They fail to identify methane in HR 8799 c but tentatively identify it in HR 8799 de's spectra.  Conversely, they argue that HR 8799 c's spectrum shows evidence for ammonia, whereas d and e lack ammonia features.    In contrast, \citet{Skemer2012} find that HR 8799 cde are of roughly the same brightness in the 3.3 $\mu m$ filter probing the presence/absence of methane.     
New, high signal-to-noise data at wavelengths diagnostic of HR 8799's atmospheric chemistry are needed to further clarify if/how HR 8799 bcde atmospheres differ.

New data may also help clarify whether our inventory of super-jovian mass planets orbiting HR 8799 is complete.   Dynamical simulations by \citet{Gozdziewski2014} suggest that an additional, fifth planet (``HR 8799 f") could be located at smaller separations.  While HR 8799 e is located close to the inner warm belt, the exact outer location of the belt is uncertain and the 2:1 resonance position with HR 8799 e may place a fifth planet slightly exterior to it.  Additionally, the truncation of the warm belt at $\approx$ 6--10 $AU$ likely requires planet-mass bodies at slightly smaller separations.   Current LBTI  $H$-band detection limits preclude the existence of another 7 $M_{J}$ companion (``HR 8799 f") exterior to $r_{proj}$ $\sim$ 0\farcs{}24 (9.5 $AU$) but  rapidly lose sensitivity at smaller separations \citep{Skemer2012}.  Sparse aperture masking observations (SAM) are, so far, only able to rule out companions more massive than 11--12 $M_{J}$ at $r$ $<$ 10 $AU$ \citep{Hinkley2011}.  

Detecting HR 8799 e with conventional AO systems like those available on Keck and Subaru, not ``extreme" AO systems like the LBT's,  requires 
employing advanced PSF subtraction methods \citep[e.g.][]{Lafreniere2007,Marois2010a,Marois2014,Currie2011a,Currie2012a}; reaching contrast limits characteristic of massive planets (e.g. 5-13 $M_{J}$) at smaller separations is even more challenging.  However, the thermal infrared at 3.5--4 $\mu m$ may present the most favorable wavelength range to search for inner planets, as the achievable Strehl ratios are higher and the planet-to-star contrasts more favorable.   Even compared to some extreme AO-assisted observations, $L^\prime$ on a conventional system coupled with advanced PSF subtraction methods may sometimes yield deeper detection limits \citep[e.g. for HD 95086][]{Rameau2013,Galicher2014}.

In this paper, we present new/archival, deep $L^\prime$ and [4.05] imaging observations of the HR 8799 system obtained with the Keck, VLT, and Subaru telescopes.  Using advanced image processing methods, we detect all four planets in each data set.  We use these data to search for a fifth planet located interior to HR 8799 e, compare the planets' atmospheric properties to those of other substellar objects, and investigate whether/how HR 8799 bcde's photometry show differences that may be due to variations in the planets' clouds or atmospheric chemistry.
 
\section{Data}
Table \ref{hr8799log} summarizes our observations which we describe in detail below.

\subsection{2012 Subaru/IRCS Data ($L^\prime$)}

We targeted HR 8799 in $L^\prime$ on 26 July 2012 with the IRCS camera \citep{Tokunaga1998} using the narrow (20.43 mas/pixel) camera
using \textit{angular differential imaging}/pupil tracking mode  \citep{Marois2006}  and without a focal plane mask (Subaru Program S12A-051).  
We modified the IRCS software to record out 0.15 $s$ short 
exposures and combine frames into data cubes every 200 frames for a cumulative integration time of 30 $s$.   
 We ``stared" at the star at a fixed position for four 10-20 minute sequences bracketed by a series of 5-10 dithered sky frames, sacrificing 
raw background-limited sensitivity for better PSF stability.   In total, we recorded just over 48 minutes of data obtained with excellent 
field rotation (173$^{o}$) in photometric conditions with average to above-average seeing.    As the stellar PSF core was saturated in each science image, 
we obtained photometric calibration data before and after our sequence.

\subsection{2012 VLT/NaCo Data ($L^\prime$, and [4.05])}
We observed HR 8799  in $L^\prime$/[4.05] on 25-26 August 2012 
with VLT/NaCo \citep{Rousset2003,Lenzen2003} in \textit{angular differential imaging} mode
without a focal plane mask with the L27 (0\farcs{}027) camera (ESO Program 089.C-0528).   For each data set, the 
exposures consist of 100-200 very short (0.1-0.5 $s$) co-added frames recorded in data cubes and taken in a four-point dither pattern.   
Our observations cover 2.5--3 hours of integration time with a field rotation of 56-73$^{o}$.   

The $L^\prime$ data were taken under light cirrus turning to photometric conditions but with a turbulent atmosphere.  
The [4.05] data suffered periodic extinction from cirrus near the end of the sequence but benefited from a more stable atmosphere.  
For the $L^\prime$ observations, we achieved flux calibration by imaging HR 8799 through a neutral density filter.  HR 8799's PSF core was unsaturated in the [4.05] data so we obtained no additional calibration images.

\subsection{2012 Keck/NIRC2 Data ($L^\prime$)}
We imaged HR 8799 on 30 October 2012 with the narrow camera 
\citep[9.952 mas/pixel;][]{Yelda2010} using "large hex" pupil plane mask but without a focal plane mask and 
in ``vertical angle" or \textit{angular differential imaging} mode (Program N087N2).   We followed our IRCS observing strategy, dithering the telescope to 
obtain sky frames three times (once at transit, twice 0.5 hours from transit) but otherwise staring on target.
Our science frames consisted of 172 21-second exposures ($t_{int}$ = 0.3 $s$, 70 coadds) for a total integration time of 3612 $s$, 
covering a field rotation of 171.1$^{o}$.   Conditions were photometric throughout the night 
with average to above-average seeing.   As the PSF core was saturated in each science image, we obtained shorter, unsaturated images of HR 8799  before and after our science sequence.

\subsection{Archival, 2010 Keck/NIRC2 Data ($L^\prime$)}
Finally, we downloaded public Keck/NIRC2 $L^\prime$ data for HR 8799 taken on 21 July 2010, which were reported in the discovery paper for HR 8799 e \citep[][shown in their Figure 1a]{Marois2010a}.   The star was positioned behind the 0\farcs{}4 diameter focal plane mask for a series of coadded exposures of 25 $s$ each bracketed by dithered sky frames.   We use $\sim$ 26 minutes worth of science frames covering a field rotation of $\sim$ 155$^{o}$.   Short, unocculted images of HR 8799 provided flux calibration.  Because the (occulted) HR 8799 PSF core is unsaturated and visible through the coronagraphic mask, we normalized the throughput of all science frames to those obtained closest in time to the unsaturated images.   As the rms of the initial star positions behind the mask was low ($\Delta$x,y $\sim$ 0.3,0.5 pixels), our flux renormalization is safe against throughput variations between the mask center and edge.
\section{Data Reduction}
\subsection{Basic Processing}
For subsequent basic processing, we followed steps outlined in \citet{Currie2010,Currie2011a,Currie2014a}, sky-subtracting a given frame using the nearest (in time) 
available sky frames and correcting each image for bad pixels.  For the NIRC2 data, we applied the non-linearity correction from \citet{Metchev2009} prior to these steps; after sky subtraction we applied distortion solutions to the NIRC2 and IRCS data.   For the data obtained in cube mode (all NaCo data, IRCS $L^\prime$), we improved PSF stability by realigning each coadd within the cube, removing frames with core-to-halo brightness ratios $\textit{{R}}$ lower than max($\textit{{R}}$ ) $<$ 3-$\sigma_{\textit{{R}} }$, and collapsing the cube using a robust mean with a 3-$\sigma$ outlier rejection.
We registered each image to a 
common center using either the unsaturated PSF ([4.05], IRCS $L^\prime$) or unsaturated PSF wings/diffraction spider (NIRC2/NaCo $L^\prime$).   
Finally, we measured the core to halo brightness ratio  across our image sequence to assess the PSF/AO correction quality and removed poorly corrected frames.  

\subsection{Speckle Suppression/PSF Subtraction}

\subsubsection{A-LOCI: basic formalism and structure}
To extract planet detections, we primarily used the adaptive, locally-optimized combination of images (A-LOCI) approach described in \citet{Currie2012a, Currie2012b, Currie2013}, which builds upon the LOCI least-squares PSF subtraction algorithm pioneered by \citet{Lafreniere2007}.     
(A-)LOCI subtracts speckles from a target image \textit{T} over a subtraction zone
\textit{S$^{T}$} by finding a set of coefficients $\sum\limits_{k=0}^{n(K)}$ \textit{x$^{k}$} for n(K) reference images that minimize residuals over a larger optimization 
region $\textit{O}$ composed of $I$ pixels.   Here, \textbf{x} = \textbf{A}$^{-1}$\textbf{b}, where \textbf{A} is a square matrix with elements \textit{A$_{j,k}$} = $\sum\limits_{i}$ \textit{O$^{j}_{i}O^{k}_{i}$} and \textbf{b} is a column matrix with elements \textit{b$_{j}$} = $\sum\limits_{i}$ \textit{O$^{j}_{i}O^{T}_{i}$}.  
Previous A-LOCI components used here include the option of ``subtraction zone centering" \citep{Currie2012a}, using a moving pixel mask to remove pixels within the subtraction zone from the optimization zone and increase point source throughput \citep{Currie2012b}, and singular value decomposition/principle component analysis (SVD/PCA) to rewrite and then invert \textbf{A} \citep{Currie2013}  \citep[see also][]{Marois2010b}.

\subsubsection{New Modifications to A-LOCI}
In addition to our SVD cutoff, we employ ``speckle filtering"/frame selection to truncate our set of reference images by their degree of correlation with the target image.  In previous versions, we performed this by computing the cross-correlation between image pixels across our reference image set within an annular area, defining our cutoff numerically as a $corr_{lim}$ less than 1.  
Here, we modify this approach by performing speckle filtering ``on the fly" for pixels within each optimization region $\textit{O}$, sorting the reference images by their degree of correlation with the target and then truncating the set by $n_{ref,t}$ elements.

 Finally, after securing an initial detection with the above steps, we model and remove the planet signal contribution to the reference PSF following methods outlined in the SOSIE pipeline from \citet{Marois2010b}.  The planet PSF affects the determination of the 
 LOCI coefficients in each reference PSF \citep[e.g.][]{Marois2010b,Brandt2012}.  Therefore, we perturb the reference image library by a noiseless planet PSF at the positions of the planets constructed from unsaturated stellar images: \textit{O$^{R^{\prime}}$} = \textit{O$^{R}$} - $\delta\times$c$PSF_{planet}$, where $\delta$ = 1 at the planet location but zero otherwise.   The coefficient $c$ is the scaling  applied to the empirical PSF needed to null the planets' signal in each reference image containing the planet (e.g. the planets' true brightnesses modulo rotational smearing over the course of an exposure).  The covariance matrix $\textbf{A}$ and column matrix $\textbf{b}$ then transform to \textbf{A$^\prime$} and \textbf{b$^\prime$}, respectively.  Thus, the coefficients $x^{k}$ are determined from a set of planet-free reference images.   The resulting combined, processed image lacks the negative PSF footprints that flank planet positions in LOCI reductions \citep{Marois2010b}.  By construction, the planets' throughputs after applying this step are 100\%.

\subsubsection{Additional tests and Alternate Reduction Methods}
We also explored alternate methods to solve the system of linear equations \textbf{A}\textbf{x} = \textbf{b} and determine the coefficients $x^{k}$, including Cholesky decomposition, \textrm{LU} factorization, and Gauss-Seidel iteration.  Cholesky decomposition yielded small ($\sim$ 5\%) gains in $SNR$ compared to a simple, double-precision matrix inversion.  But at least for our data sets the flexibility enabled by inversion of a truncated correlation matrix using SVD/PCA yielded comparable to slightly better performance. We will further explore these methods and matrix conditioning in a future paper.

As a point of reference, we also reduced our data using LOCI \citep{Lafreniere2007} with a weak SVD cutoff (10$^{-6}$) as in \citet{Currie2013b} and the classical PSF subtraction method utilized by \citet[][``sADI"]{Marois2006}.  
We did not further consider a separate, advanced approach using the Karhunen-Lo\'eve Image Projection (KLIP) algorithm \citep{Soummer2012}.  Our previous experience suggests that LOCI and KLIP achieve comparable performance under most circumstances, while A-LOCI usually yields deeper contrasts for point sources \citep[see notes in ][]{Currie2013},  and KLIP may be preferable for extended source emission (e.g. disks).   Although the two methods are often treated as separate they are mathematically very similar: truncating the covariance matrix \textbf{A} by SVD in (A-)LOCI and performing eigenvector decomposition in KLIP are equivalent \citep{Marois2014}.  An upcoming paper will explore these issues in detail as well.

\subsection{Detections}
\begin{figure}
\centering
\centering
\includegraphics[scale=0.36,clip]{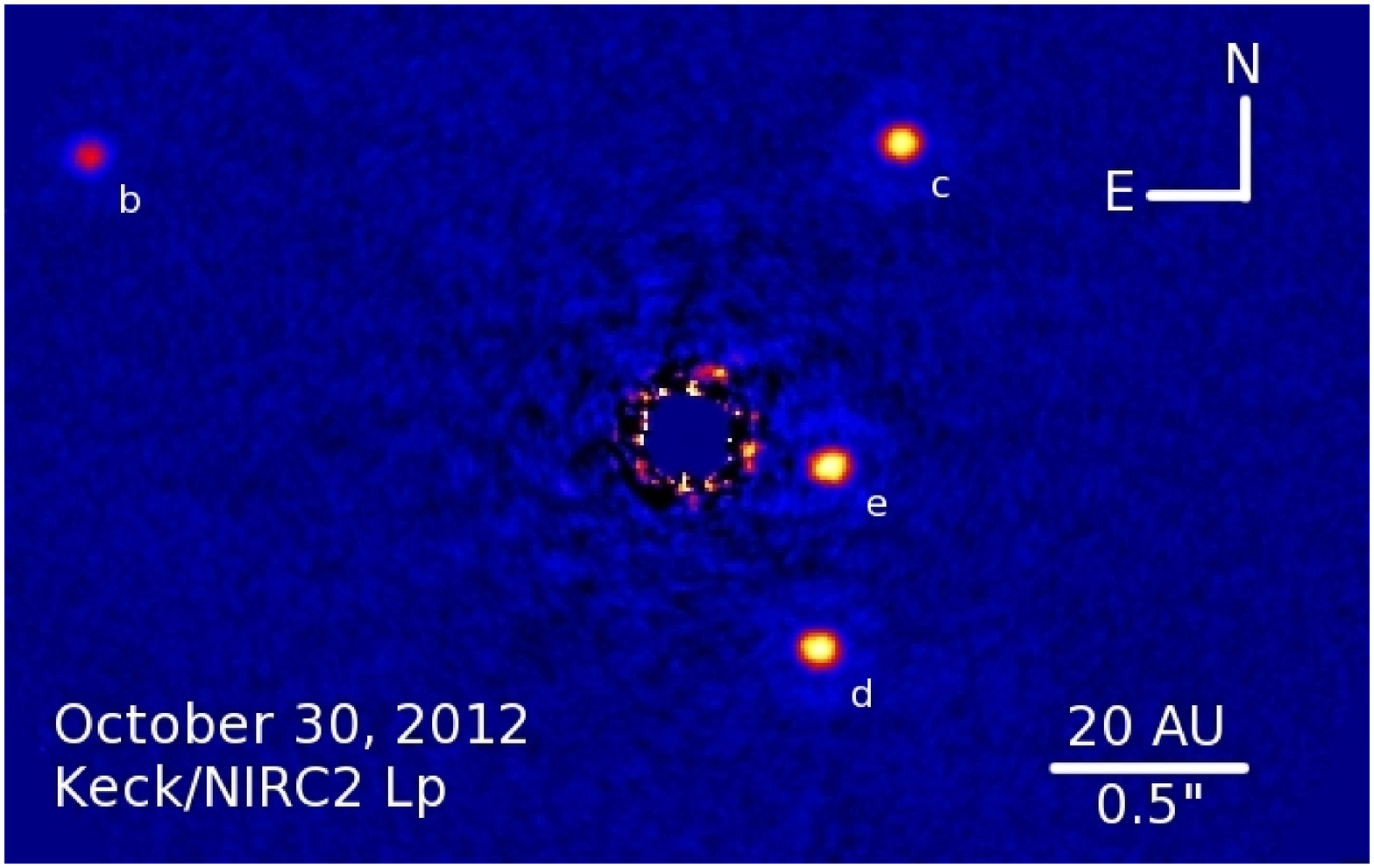}
\includegraphics[scale=0.36,clip]{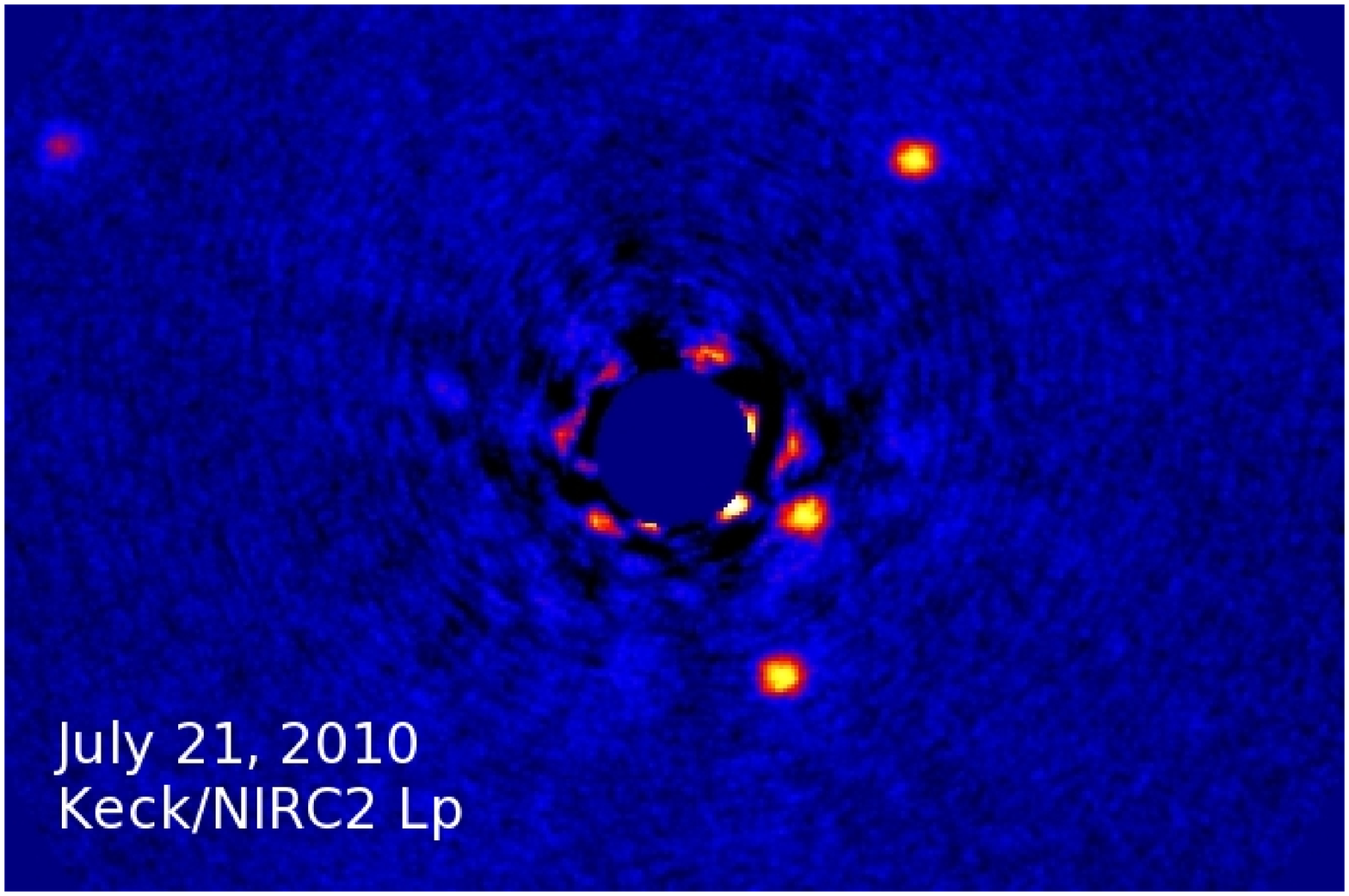}
\caption{Keck/NIRC2 $L^\prime$ images of HR 8799 from 2012 data (left) and archival 2010 data (right) reduced using A-LOCI.  HR 8799 bcde are all easily identifiable.}
\label{detimageskeck}
\end{figure}

\begin{figure}
\centering
\includegraphics[scale=0.266,trim=2mm 14mm 36mm 3mm,clip]{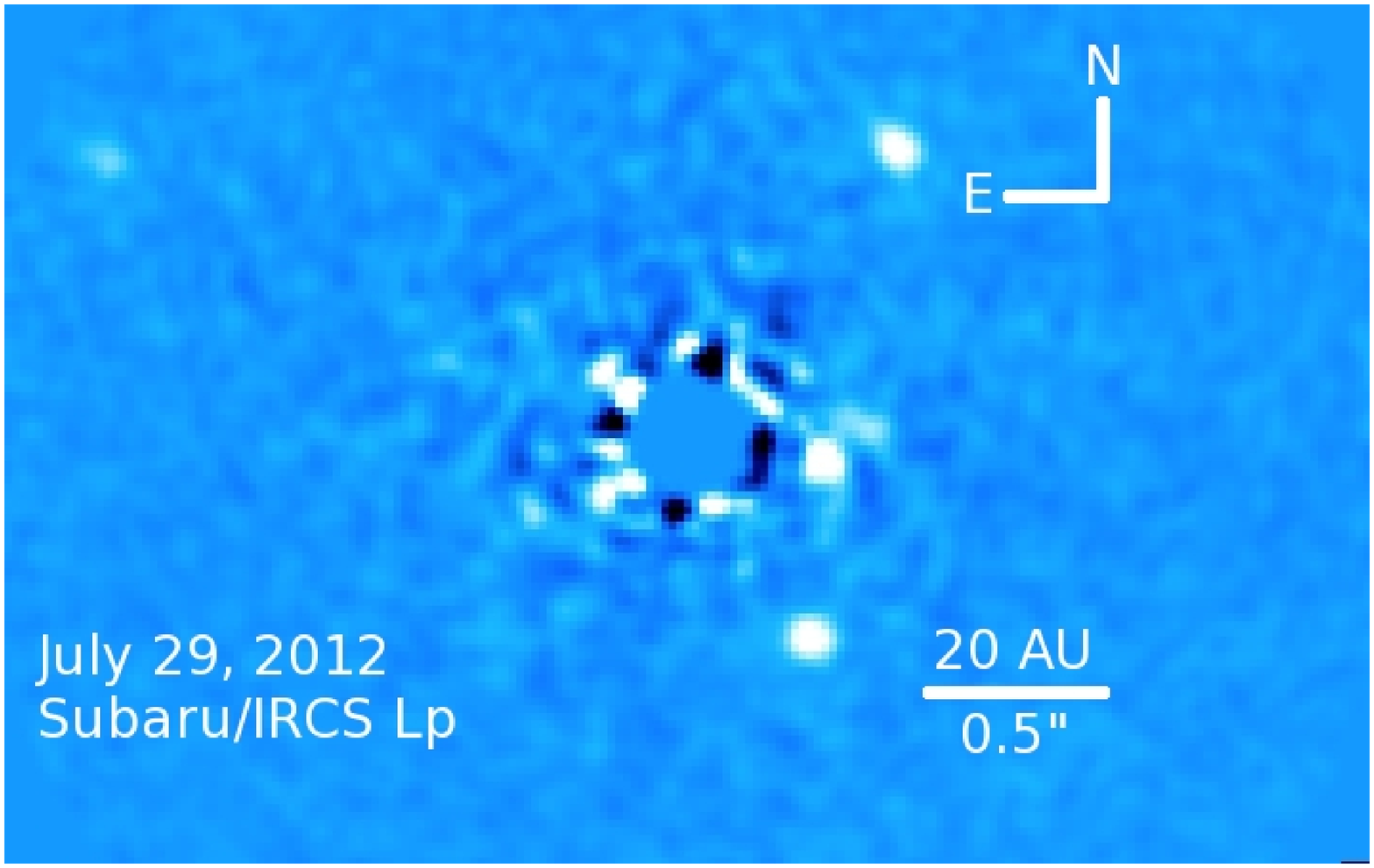}
\includegraphics[scale=0.3534,trim=35mm 30mm 57mm 22mm,clip]{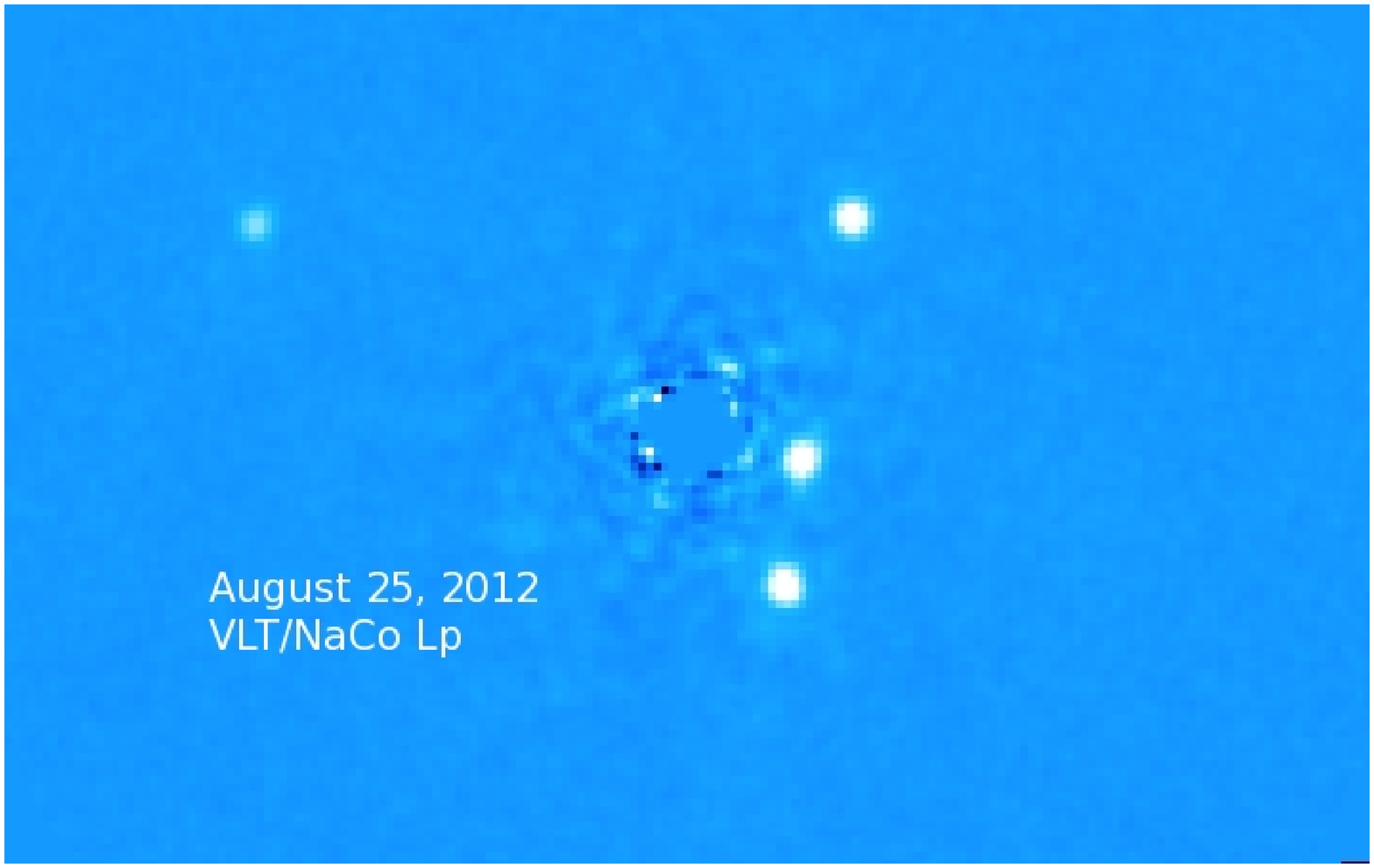}
\includegraphics[scale=0.352,trim=35mm 28mm 57mm 24mm,clip]{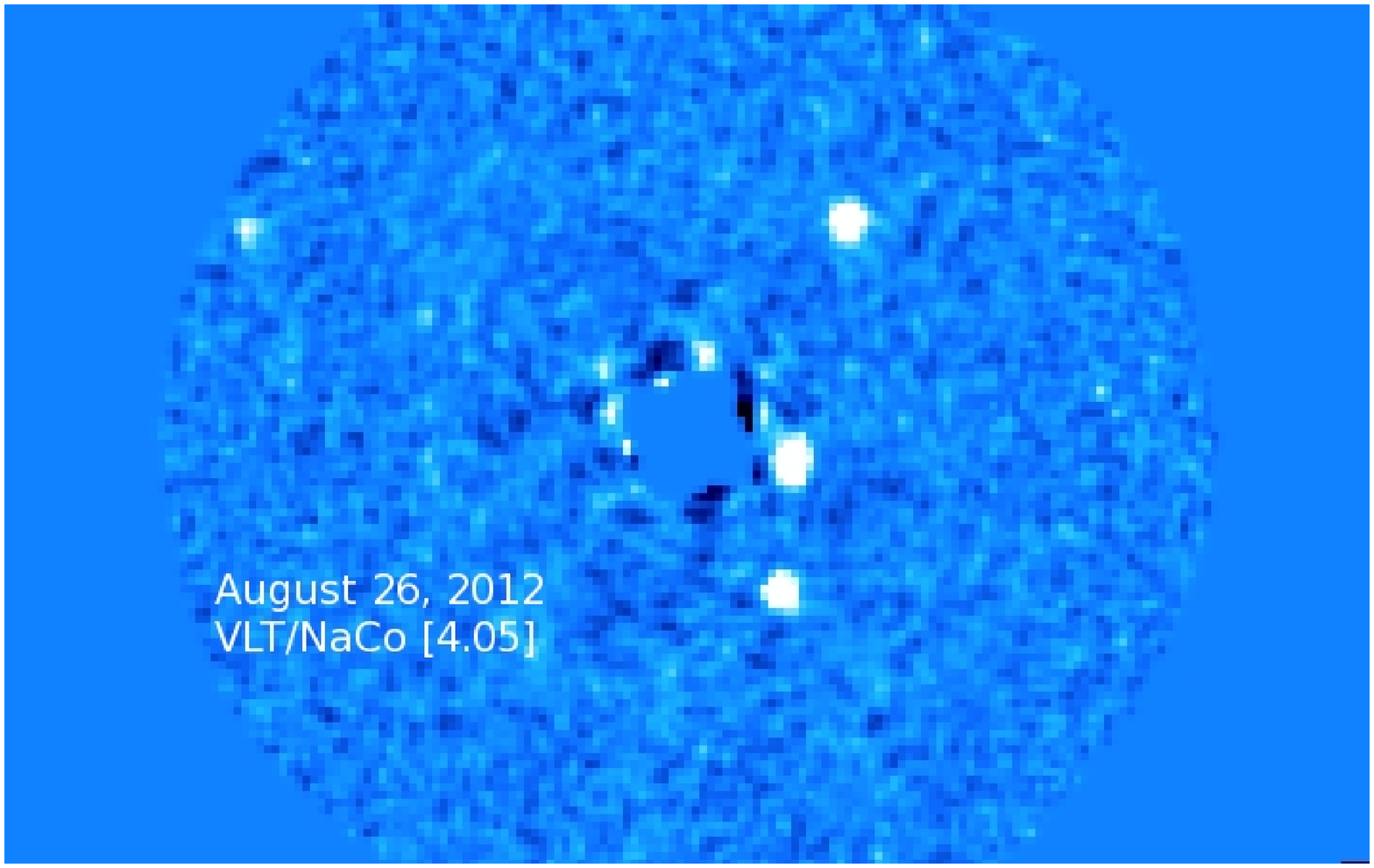}
\caption{Subaru/IRCS (left), VLT/NaCo $L^\prime$ (middle) and [4.05] (right) detections of HR 8799 bcde.  The image scales are the same in each figure panel.}
\label{detimagesvlt}
\end{figure}

Figures \ref{detimageskeck} and \ref{detimagesvlt} display our reduced Keck/NIRC2, Subaru/IRCS images, and VLT/NaCo images.   All four planets -- HR 8799 bcde -- are plainly visible in each data set.  The planets are particularly well-separated from residual speckle noise in the 2012 NIRC2 $L^\prime$ data, where the image is clean of bright residuals exterior to 0\farcs{}15--0\farcs{}2.  

To compute the SNR for HR 8799 bcde, we convolve the image with a Gaussian with an aperture of one full-width at half maximum (FWHM), compare the (convolved) flux at each planet's position to the dispersion in convolved signal at the same separation but excluding the planet \citep[e.g.][]{Thalmann2009,Currie2013}.   To be conservative, we estimate the SNR in the A-LOCI steps yielding an initial detection, \textit{without} nulling the planets' signal in the reference PSF library.

Our reductions yield SNR $\sim$ 6--15 for HR 8799 e and higher SNR for the other planets (Table \ref{hr8799phot}, middle column).    While our NaCo and IRCS $L^\prime$ detection significances for the innermost planet HR 8799 e are comparable to previous published detections, those for NIRC2 
 are significantly higher and comparable to that achieved with the (higher Strehl) LBTAO/Pisces $H$ band data presented in 
\citet{Skemer2012}  (SNR $\sim$ 10, A. Skemer pvt. comm.).    
 
For the [4.05] data, the narrow filter bandpass and high sky background limit our detections.  For all other data sets, HR 8799 b's detection is limited by photon noise whereas HR 8799 c,d, and especially e's detections are limited by our ability to suppress quasi-static speckles.   Our inner working angle for all data sets is $\sim$ 0\farcs{}13--0\farcs{}15, although only the 2012 NIRC2 data are sensitive to planet mass companions interior to 0\farcs{}25.

\begin{figure}
\centering
\includegraphics[scale=0.235,clip]{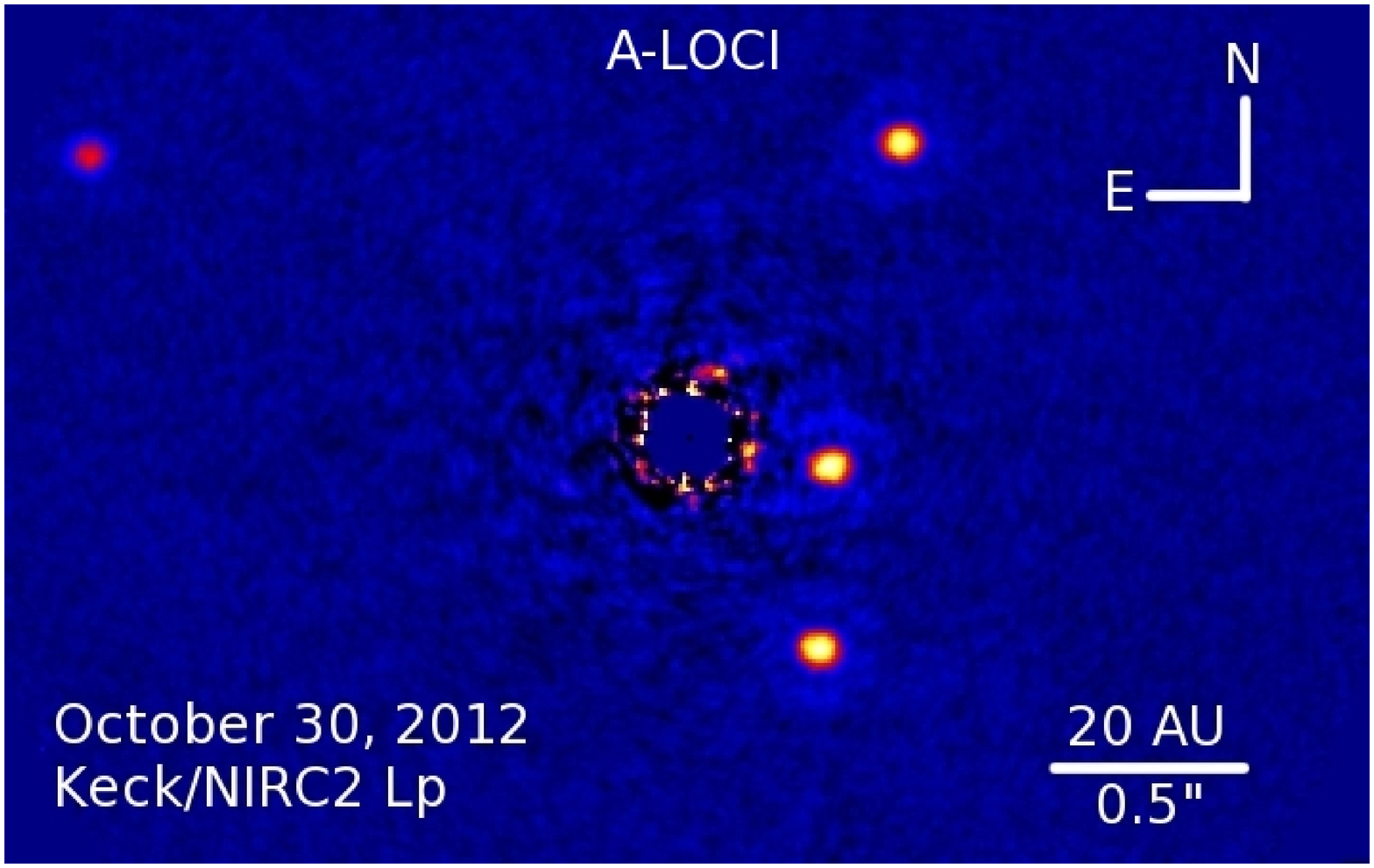}
\includegraphics[scale=0.235,clip]{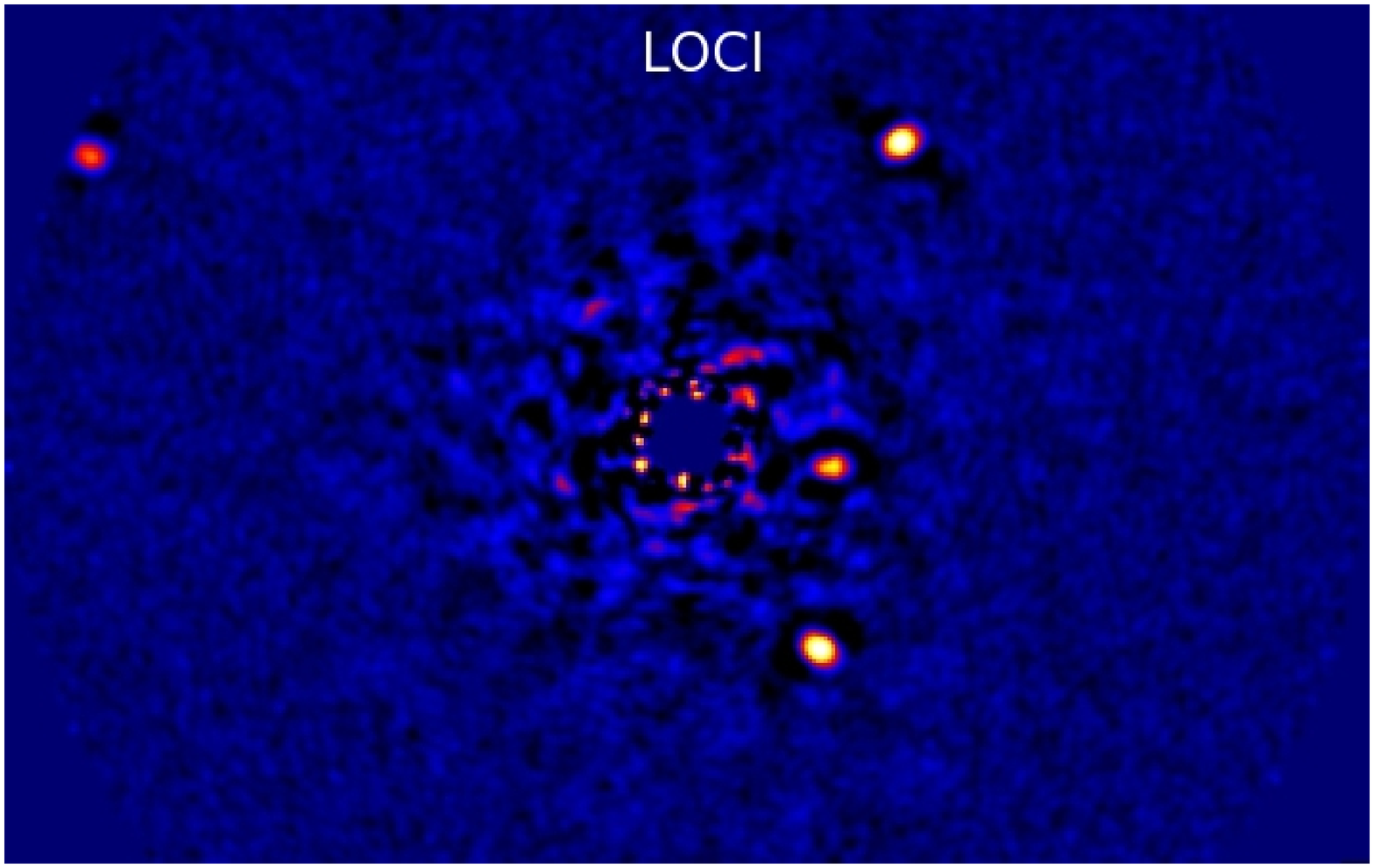}
\includegraphics[scale=0.235,clip]{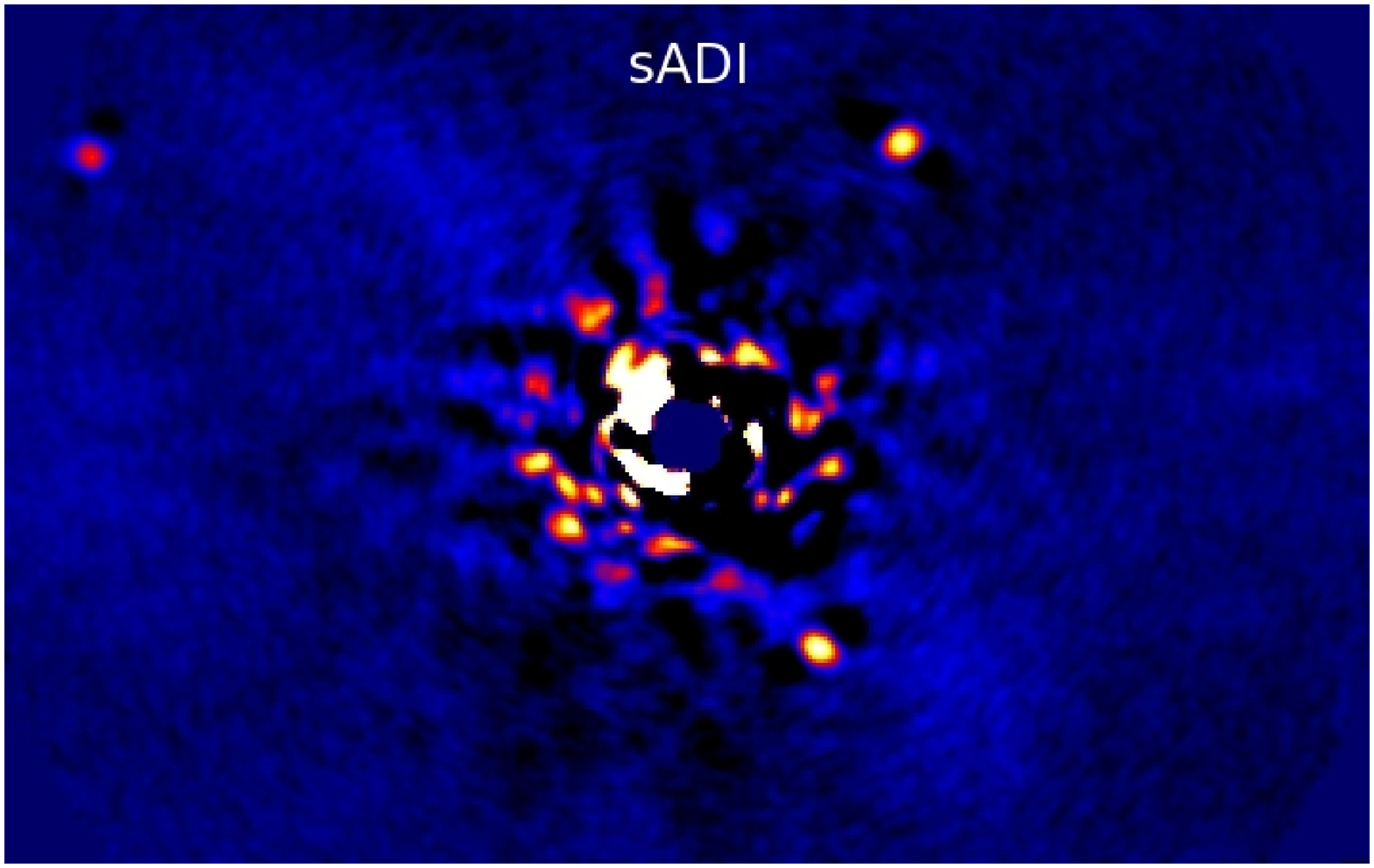}
\caption{Comparing our reduced Keck/NIRC2 $L^\prime$ image using A-LOCI (left), LOCI (coupled with an SVD cutoff; middle), and sADI (i.e. classical, ADI-based PSF subtraction; right).  While A-LOCI yields the highest throughput, highest signal-to-noise detections of HR 8799 bcde, LOCI likewise yields strong detections of all four planets (SNR $>$ 9).  HR 8799 e is undetectable at a statistically significant level using sADI.  }
\label{redcomp}
\end{figure}

The detectability of HR 8799 bcde vary with the speckle suppression method used (Figure \ref{redcomp}).
Compared to a more standard LOCI speckle suppression, we achieve a 50-100\% gain in SNR for HR 8799 e, small to negligible $SNR$ gains (0-20\%) for the other planets, but higher throughput for all planets.  These SNR and throughput gains are due to A-LOCI's more aggressive SVD cutoff that reduces errors propagating through the covariance matrix inversion, reference image set truncation (likewise reducing propagated errors and forcing the algorithm towards a more optimal, lower residual solution), and moving pixel mask forcing the algorithm to favor removing speckles instead of the planet signal \citep[see also][]{Marois2010b,Currie2012a,Currie2012b,Currie2013,Brandt2012}.  

A-LOCI and LOCI typically achieve factors of 2--3 gain in SNR compared to classical PSF subtraction for HR 8799 bcd and factors of 3-10 gain for HR 8799 e.    We fail to detect any signal from HR 8799 e at all in the IRCS data from sADI; the planet is undetectable at SNR $>$ 3 with classical PSF subtraction in all but the [4.05] data set (SNR$_{e, sADI}$ = 3.5).  

Our results strongly disagree with the NICI campaign's claims, who argue that classical ADI-based PSF subtraction performs just as well as LOCI for Strehl ratios higher than those used to first test LOCI's performance (e.g. S.R. $\gtrsim$ 40\% for NICI) \citep{Wahhaj2013} due to increased PSF stability.   The PSF quality here is even higher than for NICI (S.R. $\sim$ 70--80\%).
A future paper will explore and explain the sensitivity gains from various PSF subtraction methods at moderate to high Strehl ratios (Cloutier \& Currie 2014 in prep).

\subsection{Planet Photometry and Astrometry}
To derive photometry for HR 8799 bcde, we measured the planets' brightnesses within apertures equal to the image FWHM in each case \citep[see ][]{Currie2013}.  We assessed and corrected for planet throughput losses due to processing by two ways.  First, we compared the flux and position of synthetic point sources of known integrated flux before and after processing \citep[e.g.][]{Lafreniere2007,Currie2013}.  Second,  we used the steps outlined in \S 3.2.2 (second paragraph) to determine the brightness and position of negative copies of the unsaturated stellar PSF  that minimized the subtraction residuals.  The true planet signals estimated by these methods typically agree to within $\sim$ 5\%; similarly, the astrometry derived from both methods show excellent agreement,  within fractions of a pixel.  

For photometric calibration, we relied on unsaturated images of the primary star either bracketing the observing sequence (Keck/NIRC2 $L^\prime$) or during it (VLT/NaCo [4.05]), images of HR 8799 obtained through the neutral density filter (VLT/NaCo $L^\prime$), or separate photometric standard star observations (Subaru/IRCS).      
Following \citet{Marois2008}, we assumed an apparent magnitude of 
m$_{L^\prime}$ = 5.22 $\pm$ 0.02 for HR 8799.   We adopted an identical magnitude at [4.05]: m$_{[4.05]}$ = 5.22 $\pm$ 0.02.   To derive photometric uncertainties for each measurement, we considered the S/N of our detection, the uncertainty in the planet throughput, and the uncertainty in absolute flux calibration \citep[see][]{Currie2013}.

The right two columns of Table \ref{hr8799phot} report our derived planet-to-star contrasts ($\Delta$m$_{L^\prime, [4.05]}$) and absolute magnitudes along with uncertainties in these measurements.    The $L^\prime$ photometry derived from different sources are consistent with one another within photometric errors.   While the relative sources of photometric error vary, typically the throughput uncertainty dominates the error budget \citep[see also][]{Currie2013}.  

Comparing the $L^\prime$ photometry from each source and to previously published measurements by \citet{Marois2008} and \citet{Currie2011a}, the 2012 Keck photometry appear to be systematically brighter by $\approx$ 0.05 mag.   While this discrepancy could in principle be due to light cirrus during the photometric standard frames, detector non-linearity is another likely explanation.  The non-linearity correction from \citet{Metchev2009} was derived from flat-field lamps with high illumination: the linearity correction for fainter objects is weaker than derived, leading us to slightly underestimate the planet-to-star contrast.    Indeed, with unsaturated images from the archival 2010 Keck/L$^\prime$ data, taken using shorter exposures (and thus deeper in the linear count regime), we derive planet photometry systematically fainter by $\sim$ 2.5-10\%.   The 2010 NIRC2 photometry also appear slightly more consistent with measurements we have obtained from archival NIRC2 data taken on September 2008 and October 2010 and reported in \citet{Marois2008,Marois2010a} (T. Currie, unpublished).  While our data yield the first detections of HR 8799 bcde in the [4.05] filter, the flux density for HR 8799 c averaged over 4--4.1 $\mu m$ reported by \citet{Janson2010} agrees with our [4.05] photometry to within errors.\\

Table \ref{hr8799astrom} lists our 2010 and 2012 Keck/NIRC2 astrometry.   For HR 8799 cde, our centroiding uncertainty (0.5 pixels along either axis) dominates the astrometric error budget; for HR 8799 b the north position angle uncertainty also becomes important.  We do not derive astrometry for the NaCo or IRCS data sets, because they require additional analyses to fine-tune their north position angles and pixel scales, although our measurements are consistent to first order.   Between the NaCo/IRCS and 2012 NIRC2 data sets HR 8799 bcde appears to exhibit slight counterclockwise motion.  This motion  is obvious between the 2010 and 2012 NIRC2 data sets.   Compared to HR 8799 bcd astrometry from \citet{Marois2008} and \citet{Currie2012b}, in 2012 HR 8799 d appears to be increasing in projected separation.   An upcoming paper will present revised orbital fits to all four planets.

\section{Limits on the Presence of a Fifth Planet, HR 8799 ``f", at Small Angular Separations}

\begin{figure}
\centering
\includegraphics[scale=0.35,clip]{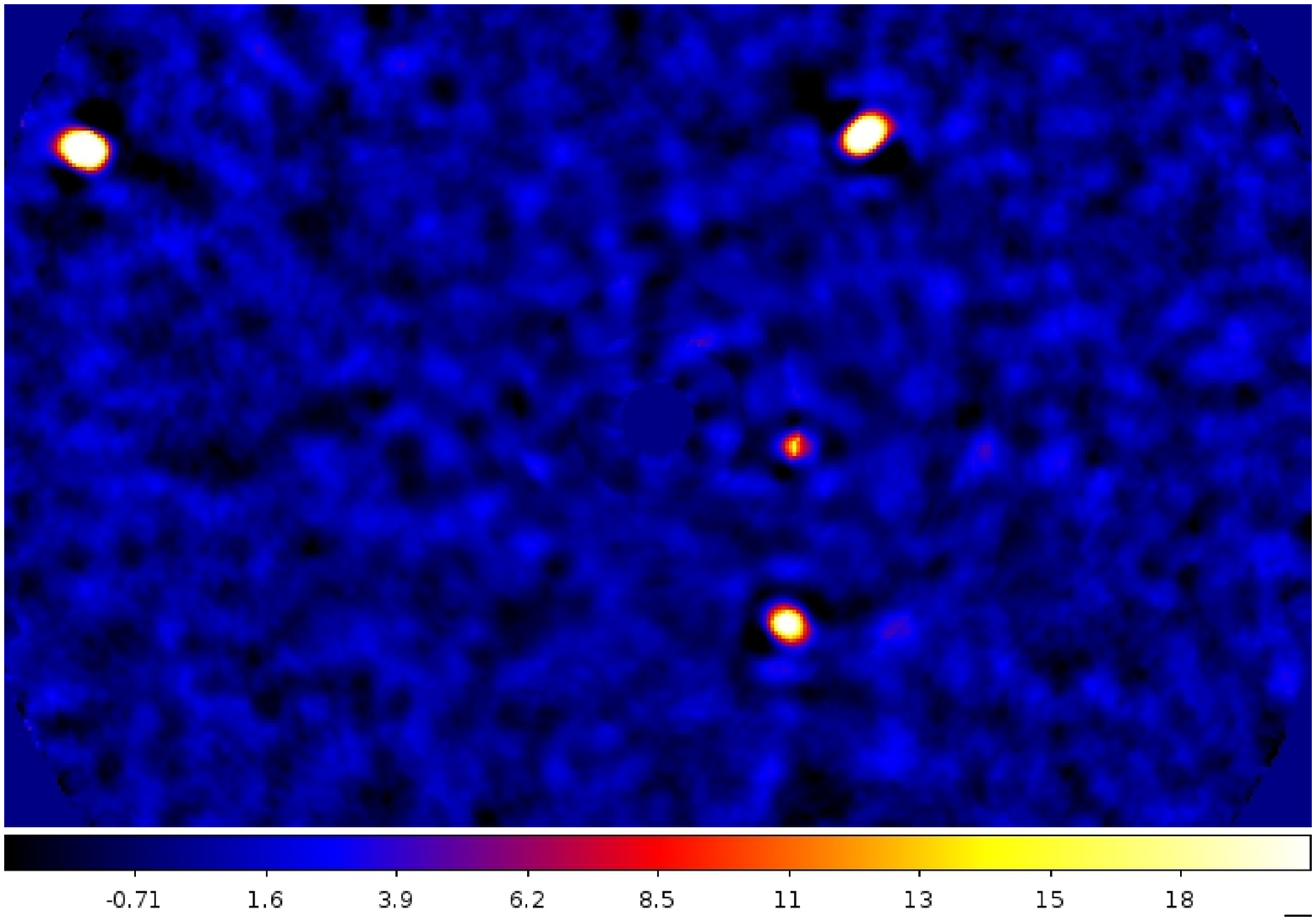}
\includegraphics[scale=0.35,clip]{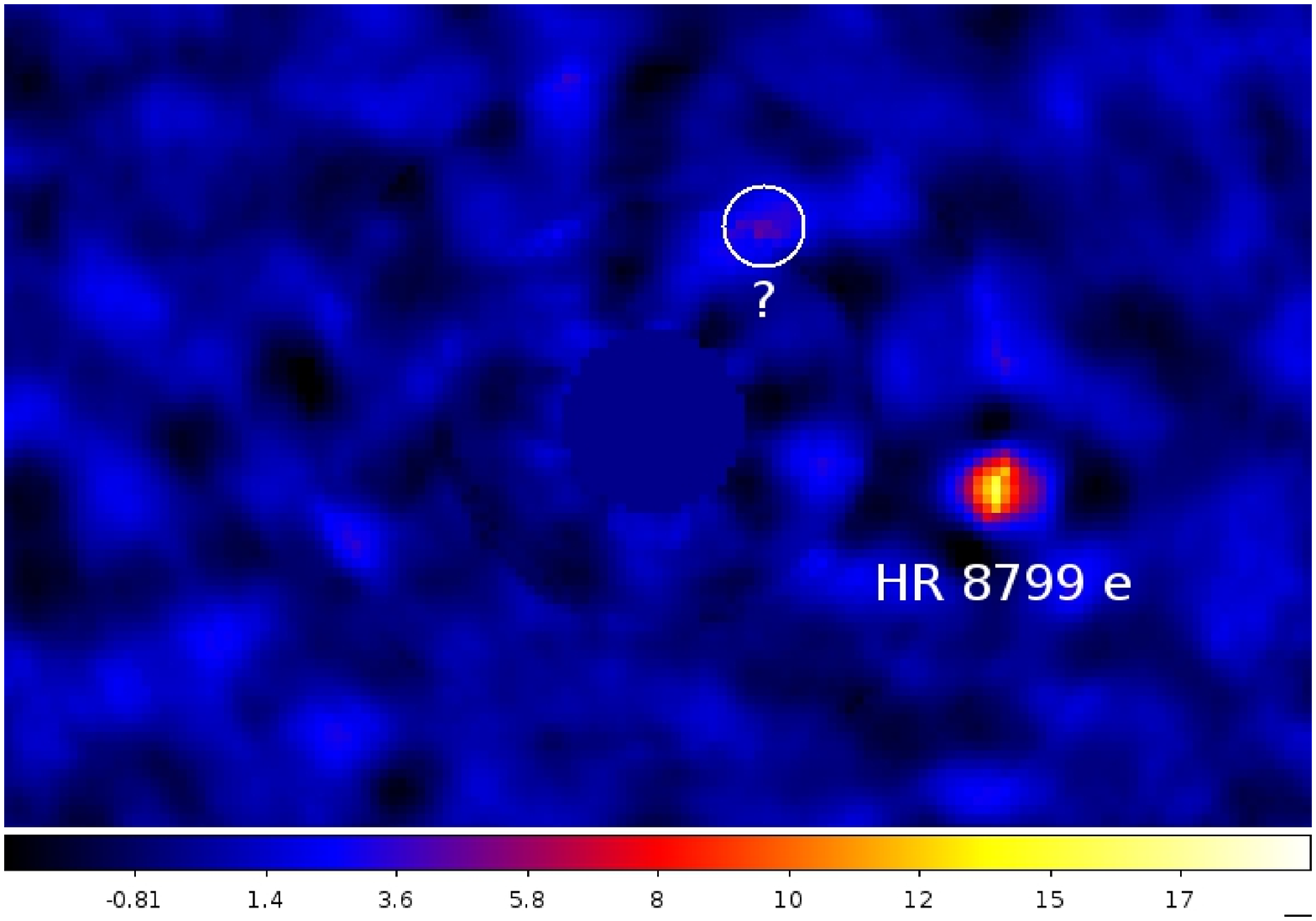}\\
\includegraphics[scale=0.5,clip]{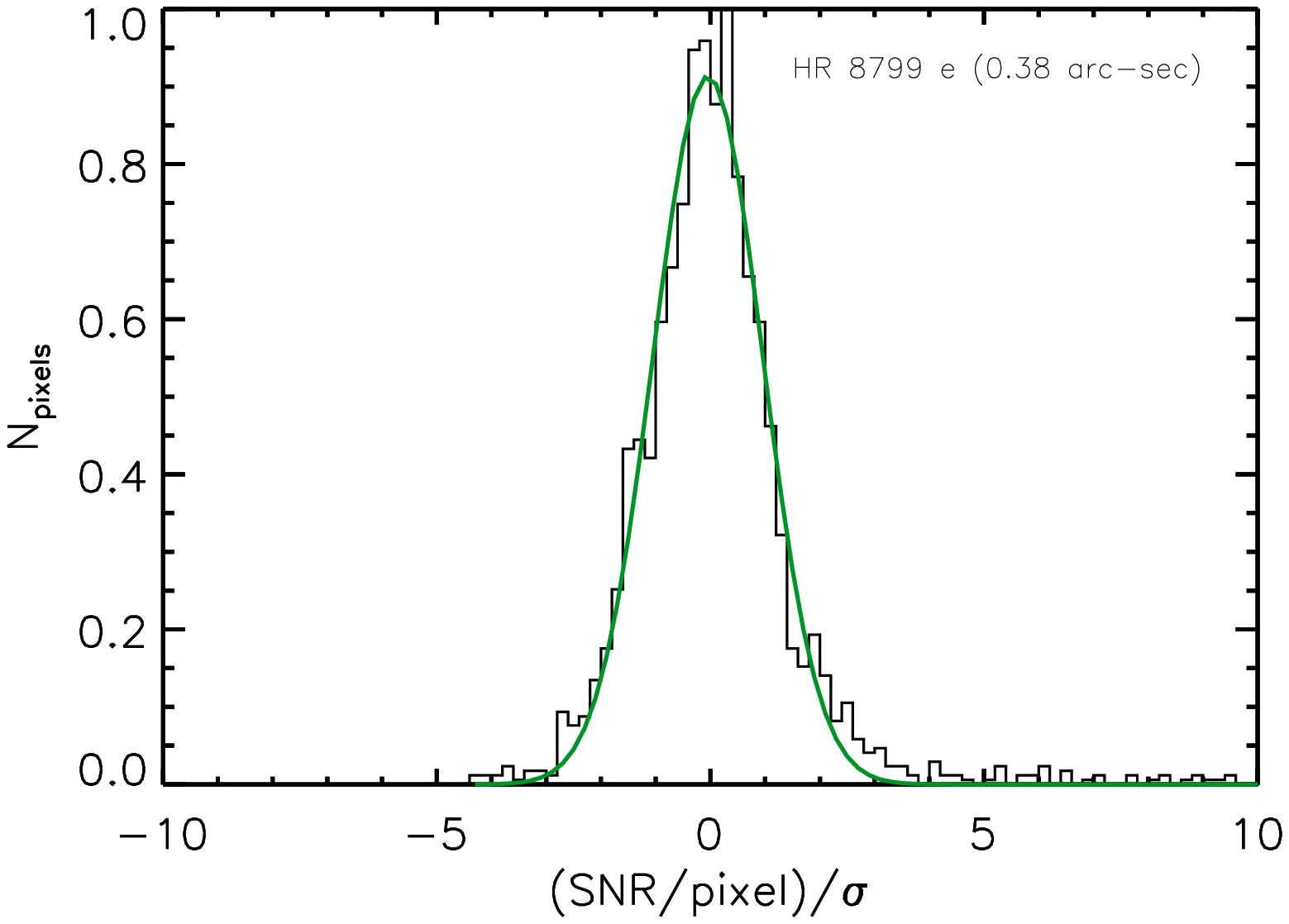}
\includegraphics[scale=0.5,clip]{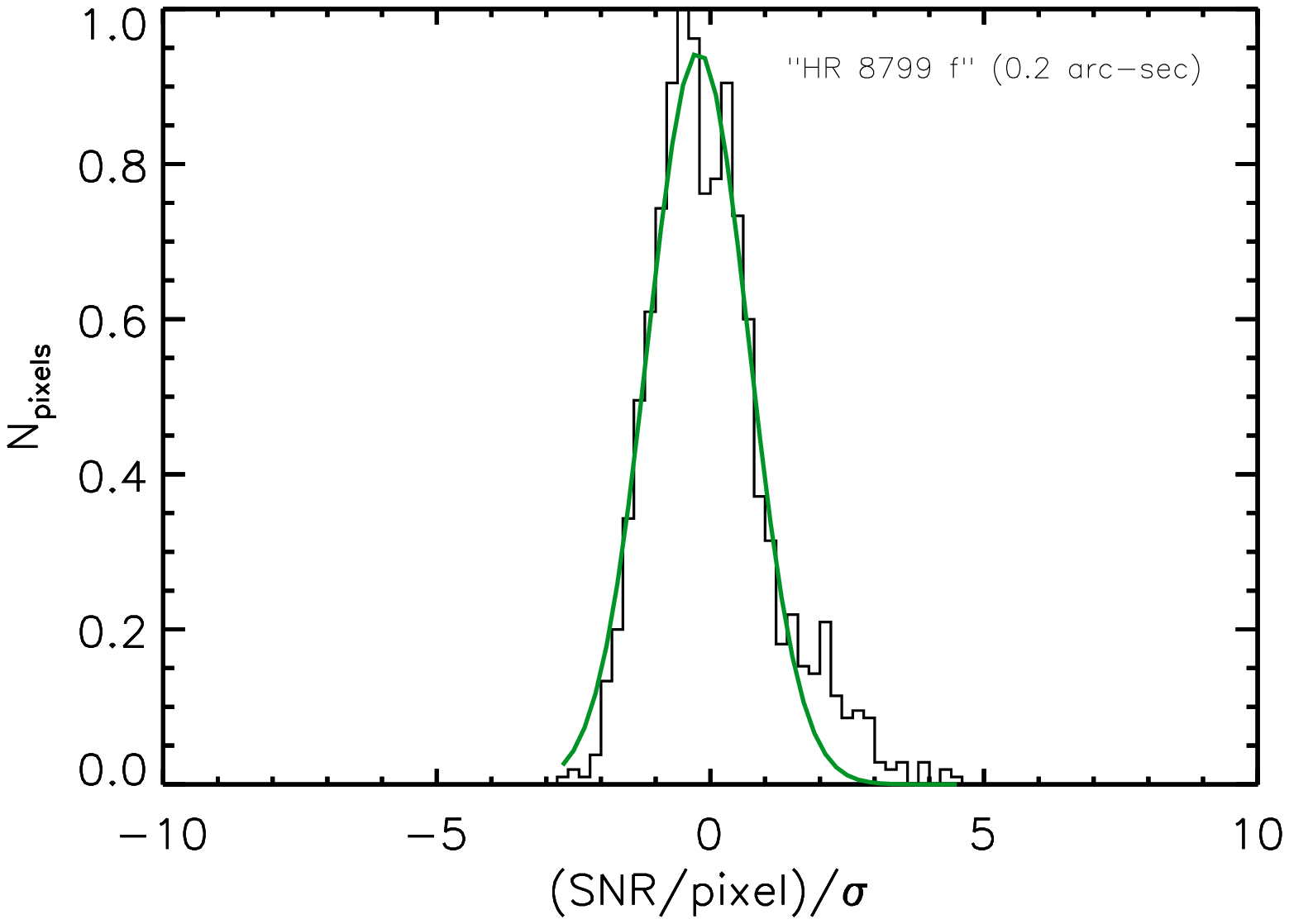}
\caption{(Top panels) Signal-to-noise ratio maps for our 2012 Keck/NIRC2 $L^\prime$ data.  The SNR is derived in our reduction steps based on an initial detection, \textit{prior} to nulling the planet signals in the reference image library.  The left panel shows the full image; the right panel shows a close-up view of the region interior to HR 8799 e with a 4-$\sigma$ peak at $r$ $\sim$ 0\farcs{}2 highlighted.  (Bottom panels) Pixel histogram distributions for our SNR maps at the location of HR 8799 e (left) and at 0\farcs{}2.  The pixel distribution at 0\farcs{}2 shows a slightly positive skew.}
\label{hr8799fdet}
\end{figure}

Figure \ref{hr8799fdet} (top panels) displays the SNR map for our most sensitive data set, the 2012 Keck/NIRC2 $L^\prime$ data.    As shown in the top-left panel (full image), HR 8799 bcde are easily identifiable as bright peaks in the map.  No other regions exhibit cluster peaks $\gtrsim$ 5-$\sigma$, consistent with there being only four detected planets at $r$ $>$ 0\farcs{}3, in agreement with multiple independent recent results \citep[e.g.][]{Marois2010a,Currie2011a,Skemer2012}.  

A closer look at the map at separations interior to HR 8799 e (top-right panel) reveals one suspicious 4 $\sigma$ peak  at $r$ $\sim$ 0\farcs{}2 separation ([-0\farcs{}1,0\farcs{}19], $a_{proj}$ $\sim$ 7.9 $AU$).   Besides the pixels identifying HR 8799 bcde, it is the most statistically significant local maximum on our SNR map.  Furthermore, the location is at roughly the same angular separation as and within $\sim$ 5-10 pixels of the predicted location of an ``HR 8799 f" from the Vd dynamical model proposed by \citet{Gozdziewski2014}, which puts HR 8799 e in a 2:1 mean-motion resonance with this hypothetical planet.  

However, while suggestive, this peak at $r$ $\sim$ 0\farcs{}2 is far more likely to be a PSF subtraction artifact than a real point source.  Deep $H$-band imaging from \citet{Skemer2012} fail to identify any peak residuals at slightly larger separations.  We fail to identify it in any other HR 8799 data set.  While \citet{Gozdziewski2014} posit the Vd dynamical setup with 5 HR 8799 planets, this configuration is disfavored compared to other, more stable models with ``HR 8799 f" located elsewhere.  

Most importantly, while the noise statistics at the location of HR 8799 e are gaussian (Figure \ref{hr8799fdet}, bottom-left panel), the residuals $r$ $\sim$ 0\farcs{}2 are slightly skewed towards positive values (Figure \ref{hr8799fdet}, bottom-right panel).  Thus, a "4-$\sigma$" detection in our map does not directly translate into a 4-$\sigma$ confidence limit in gaussian statistics \citep[see discussion in][]{Marois2008b}.  A 5+ $\sigma$ detection would be required to cleanly distinguish between a real point source and residual speckle noise.

\begin{figure}
\includegraphics[scale=0.5]{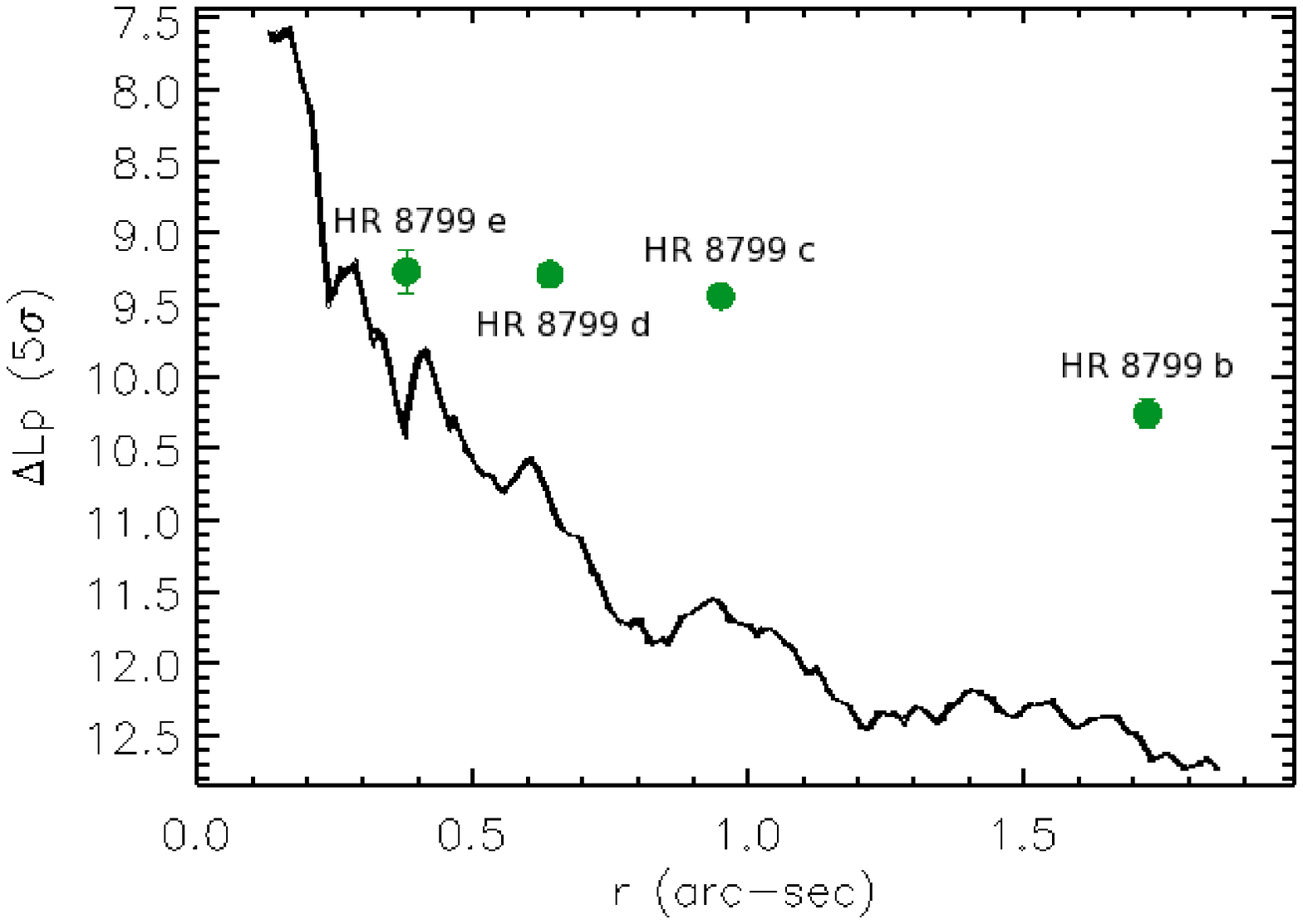}
\includegraphics[scale=0.44]{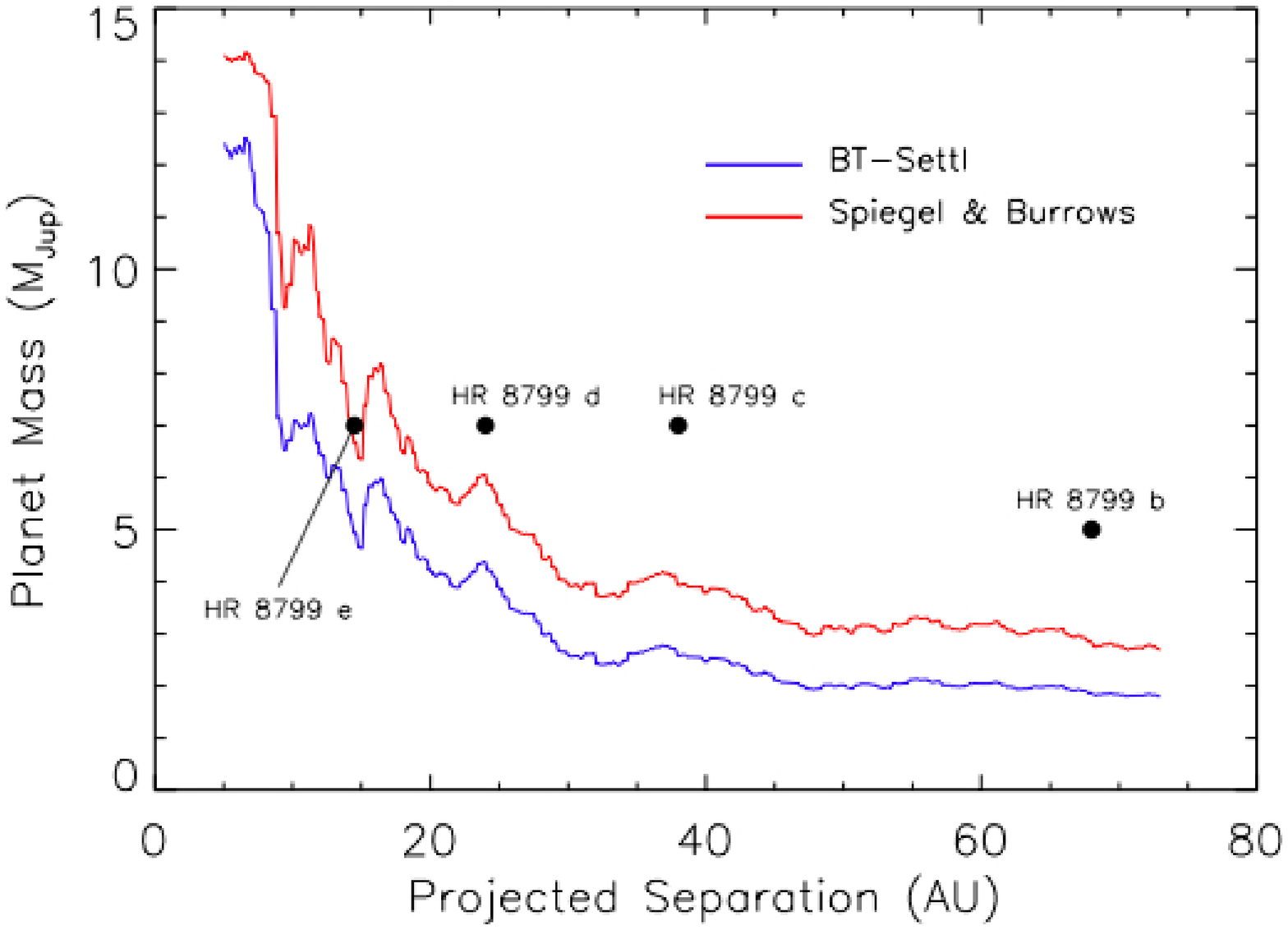}
\caption{(Left) The planet-to-star contrast for our NIRC2 $L^\prime$ as a function of angular separation.  The contrast curve exhibits a ``sawtooth"-like pattern  due to the NIRC2 PSF structure and variable planet throughput within each subtraction zone \citep[see ][]{Marois2010b}.   (Right) The inferred mass detection limits assuming planet ages of 30 $Myr$ and the \citet{Baraffe2003} luminosity evolution models + the BT-Settl atmosphere models (blue curve) and the \citet{Spiegel2012} luminosity evolution/atmosphere models (red).}
\label{contrast}
\end{figure}

We use our most sensitive data set (2012 Keck/NIRC2 $L^\prime$ data set) to derive a contrast curve and place limits on the presence of a fifth planet located interior to HR 8799 e, using the primary star for flux calibration, synthetic point sources to calibrate the point-source throughput vs. angular separation and our Gaussian-convolved image to measure $\sigma_{r}$ (Figure \ref{contrast})\footnote{As noted recently by \citet{Mawet2014} any contrast curve is sensitive to the noise properties (e.g. Gaussian, Rician) and is limited by the number of resolution elements available from which to estimate this noise and derive contrast.   They argue that the latter is usually neglected.   We do not believe this is an issue for our data.  First, the Keck $L^\prime$ PSF is very oversampled, containing many resolution elements at 1--3 $\lambda$/D.  Additionally, how $\sigma$ in a SNR calculation is determined is critical, and our method is conservative.    Based on visual inspection of their Figure 3, their injected 5-$\sigma$ planet at 1.5 $\lambda$/D has pixel intensities barely larger than other residuals at the 7 o'clock and 10 o'clock positions.   Our IRCS detection of HR 8799 e, barely over 5-$\sigma$, appeared to stand out from the residual noise far more clearly.  Based on our experience, we would likely find their injected companion to be a 1--2 $\sigma$ perturbation, not a 5-$\sigma$ detection.}.  The NIRC2 data reach contrasts at 1--1.5\arcsec{} of $\sim$ 10$^{-5}$ -- typical of those routinely achieved in the near-IR with conventional AO systems \citep[e.g.][]{Lafreniere2007b} --  $\sim$ 10$^{-4}$ at $r$ $\sim$ 0\farcs{}35, and $\sim$ 10$^{-3}$ at our inner working angle of $\approx$ 0\farcs{}13.    Keck employs a ``conventional" adaptive optics system.  Nevertheless, the advantageous observing setup for HR 8799 (excellent field rotation, stable atmosphere) and advanced PSF subtraction techniques yield contrasts at $r$ $<$ 0\farcs{}5 comparable to or slightly better than preliminary, characteristic ones reported by \citet{Skemer2014} for LBT/LMIRCam, which benefits from a much higher-actuator, \textit{extreme-AO} system capable of yielding much higher Strehl ratios\footnote{ On the other hand, their limits at wide separations are far superior owing to the low-emission, adaptive secondary design of LBTAO.  Further upgrades (i.e. advanced coronography) will improve their contrasts well over what is achievable in this work with Keck at small, HR 8799 e-like separations.}.  

From Figure \ref{contrast} (left panel), our NIRC2 data rule out the presence of any HR 8799 cde-luminosity planets at separations greater than $r$ $\sim$ 0\farcs{}2, although HR 8799 b-like companions would remain undetectable at the 5-$\sigma$ level interior to $\sim$ 0\farcs{}3.    At separations of $r$ $\sim$ 0\farcs{}13-- 0\farcs{}2, a planet with a $\beta$ Pic b-like contrast  \citep[$\Delta$L$^\prime$ $\sim$ 7.8;][]{Currie2013} would be detectable.  

To convert our contrast limits into planet masses, we consider two sets of planet cooling models: hot-start models from \citet{Baraffe2003} and separate hot-start ones from \citet{Spiegel2012}.  The massive jovian planets imaged thus far may effectively be younger than their host stars by up to $\sim$ 10 $Myr$ if they formed near the end of the protoplanetary disk lifetime \citep[c.f.][]{Hernandez2007,Currie2009,Cloutier2014}.  Thus, comparing the planet's luminosity to predictions from cooling models at the star's age may yield overestimated planet masses at young ages (e.g. 1--50 $Myr$) \citep{Currie2013}.  Even these standard hot-start models may predict the masses of substellar objects (planets, brown dwarfs) larger (e.g. $\sim$ 25\%) larger than dynamical mass estimates \citep{Dupuy2009,Dupuy2014}.  Nevertheless, to be conservative we assume that the planets are of the same age as the star and their luminosity evolution is described by standard hot-start models.

The righthand panel of Figure \ref{contrast} displays our limits for a fifth, hitherto undetected planet.    Using the \citet{Baraffe2003} evolutionary models, our contrast limits rule out the existence of an HR 8799 b-like,  5 $M_{J}$ planet exterior to about 0\farcs{}3 ($\sim$ 12 $AU$).    At $\sim$ 9 $AU$, within HR 8799's warm dust belt as calculated by \citet{Su2009} and similar to the outer of the two  predicted locations for a hypothetical ``HR 8799 f" from  \citet{Gozdziewski2014}  ( $r$ $\sim$ 7.5-9.7 $AU$ or $r_{proj}$ $\sim$ 7--9 $AU$) our data still rule out an HR 8799 cde-mass companion.   Just interior to the inner edge of the predicted warm dust belt \citep[see][]{Su2009} at 5--6 $AU$, our data rule out 12 $M_{J}$ companions.  For the \citet{Spiegel2012} models, our limits systematically shift to higher masses: $\sim$ 7 $M_{J}$ at 12 $AU$, 10 $M_{J}$ at 9 $AU$, and 13 $M_{J}$ at 5--6 $AU$.  If the HR 8799 planets formed late (e.g. $t_{form}$ $\sim$ 10 $Myr$), then the mass limits derived from the \citet{Spiegel2012} models using these adjusted ages ($\sim$ 20 $Myr$) are consistent with or slightly lower than those derived from the \citet{Baraffe2003} models assuming nominal ages.

Different assumptions about planetary atmospheres may result in deeper detection limits.  In particular, models including thicker clouds and more atmospheric dust than the BT-Settl models may better reproduce simultaneous photometry and spectra for young super-jovian companions \citep{Currie2014b}.   Adopting instead the AMES-DUSTY models to map between contrast and planet mass, our mass limits are \textit{more} sensitive in some cases: 5.5 $M_{J}$ at 9 $AU$ and 4.5 $M_{J}$ at 12 $AU$.  In contrast, dustier atmospheres render super-jovian planets fainter in $H$-band and thus mass limits from platforms like GPI, SPHERE, and LBT/Pisces can be significantly \textit{less} sensitive.   Likewise, detection limits in $H$ relative to $L^\prime$ degrade if warm-start models, not hot-start models, describe the luminosity evolution of super-jovian planets \citep[c.f.][]{Spiegel2012}.

\section{Thermal IR Constraints on the Atmospheres of HR 8799 \lowercase{bcde}}
In this section, we compare HR 8799 bcde's photometry to other substellar objects, to each other, and to atmosphere models.  Table \ref{tab_litphot} lists our adopted photometry.  Here, we consider the 2010 Keck/NIRC2 $L^\prime$ data instead of the 2012 photometry because the former were computed from data more in the linear count regime.  
\subsection{Comparisons to Field L and T Dwarfs}
Combined with our updated $L^\prime$ photometry, our new [4.05] data allow us to more completely assess how HR 8799 bcde's broadband colors (and, by inference, atmospheric properties) compare to those for other imaged planetary companions and to field brown dwarfs.  For brown dwarfs, we consider the \citet{Leggett2010} sample of field L/T substellar objects targeted in $JHK_{s}L^\prime$ broadband filters and the \textit{AKARI}-selected sample of 3--5 $\mu m$ brown dwarf spectra from \citet{Sorahana2012}.   We derive photometry for the \citeauthor{Sorahana2012} sample at $L^\prime$ and [3.3] by convolving each spectrum with the appropriate \textit{Mauna Kea Observatory} filter functions.  

 We also include planetary companions $\beta$ Pic b \citep{Lagrange2010}, ROXs 42Bb \citep{Currie2014a}, and HD 95086 b \citep{Rameau2013} and other (candidate) planet-mass companions $\kappa$ And b \citep{Carson2013}, 1RXJ 1609 B \citep{Lafreniere2008}, GSC 06214 B \citep{Ireland2011}, USco CTIO108 B \citep{Bejar2008}, 2M 1207 b \citep{Chauvin2004}, and 2M0103AB B \citep{Delorme2013} \footnote{Here, we emphasize the word \textit{candidate} planet-mass companions.  Whether some companions -- e.g. $\kappa$ And b -- have inferred masses lying above or below the deuterium-burning limit depends on the companion's uncertain age \citep[e.g.][]{Hinkley2013}}.   Photometry for $\beta$ Pic b and ROXs 42Bb draw from \citet{Currie2013,Currie2014a,Currie2014b}; measurements for HD 95086 b derive from \citet{Galicher2014}.    Photometry for other companions draw from the compilation in \citet{Currie2013}.
 
\begin{figure}
\centering
\includegraphics[scale=0.52,trim=10mm 5mm 5mm 0mm,clip]{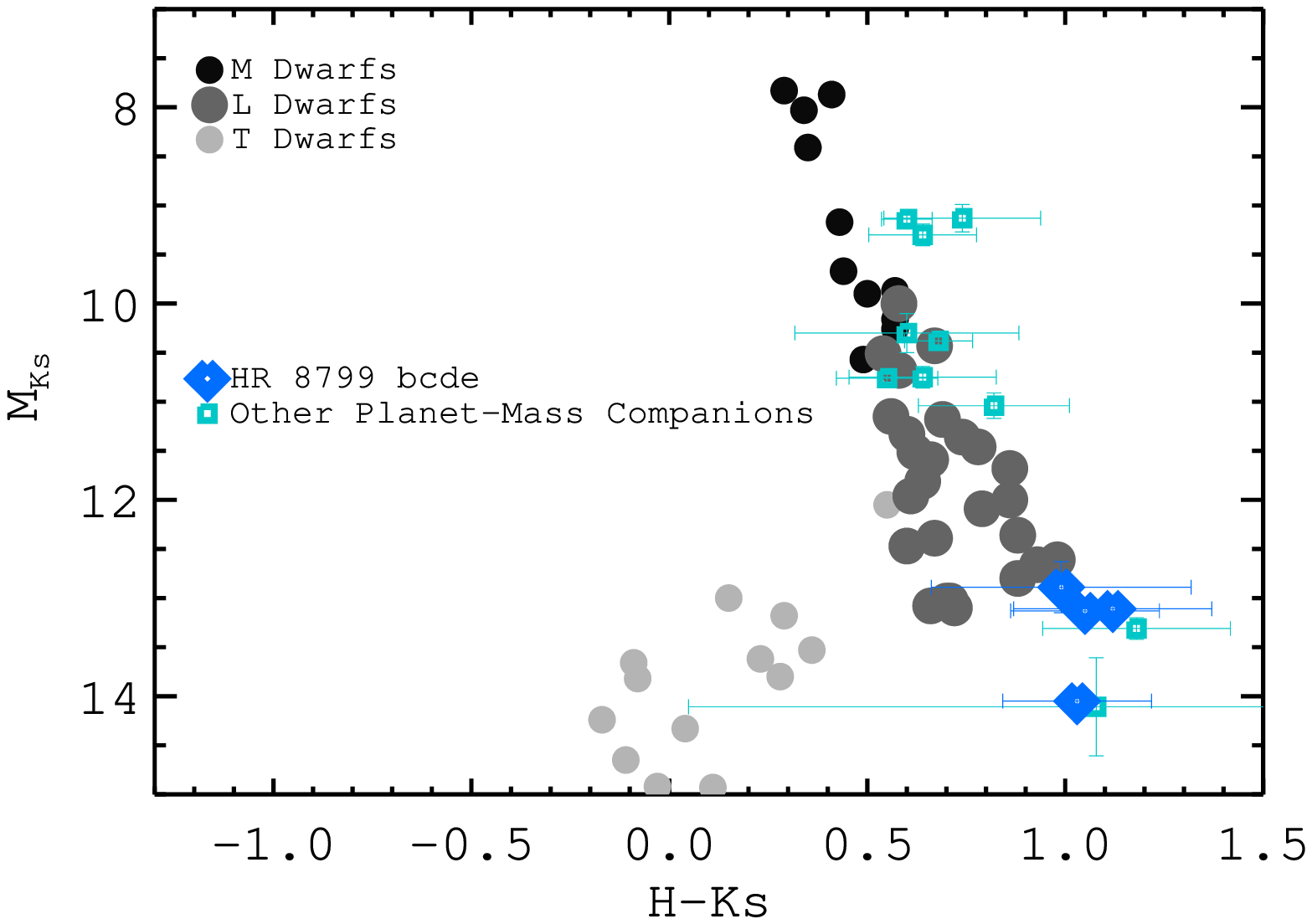}
\includegraphics[scale=0.52,trim=10mm 5mm 5mm 0mm,clip]{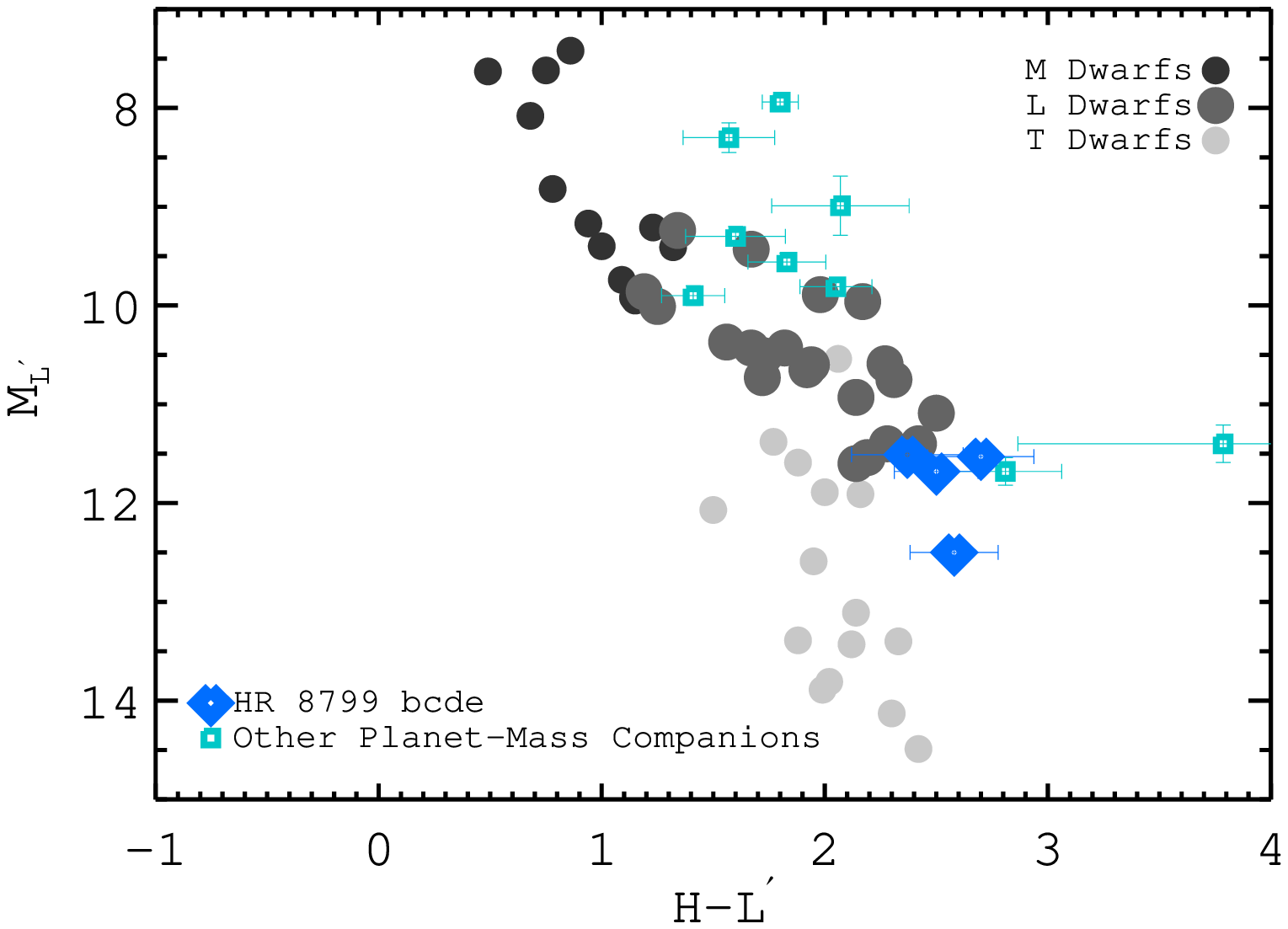}
\\
\includegraphics[scale=0.52,trim=10mm 5mm 5mm 0mm,clip]{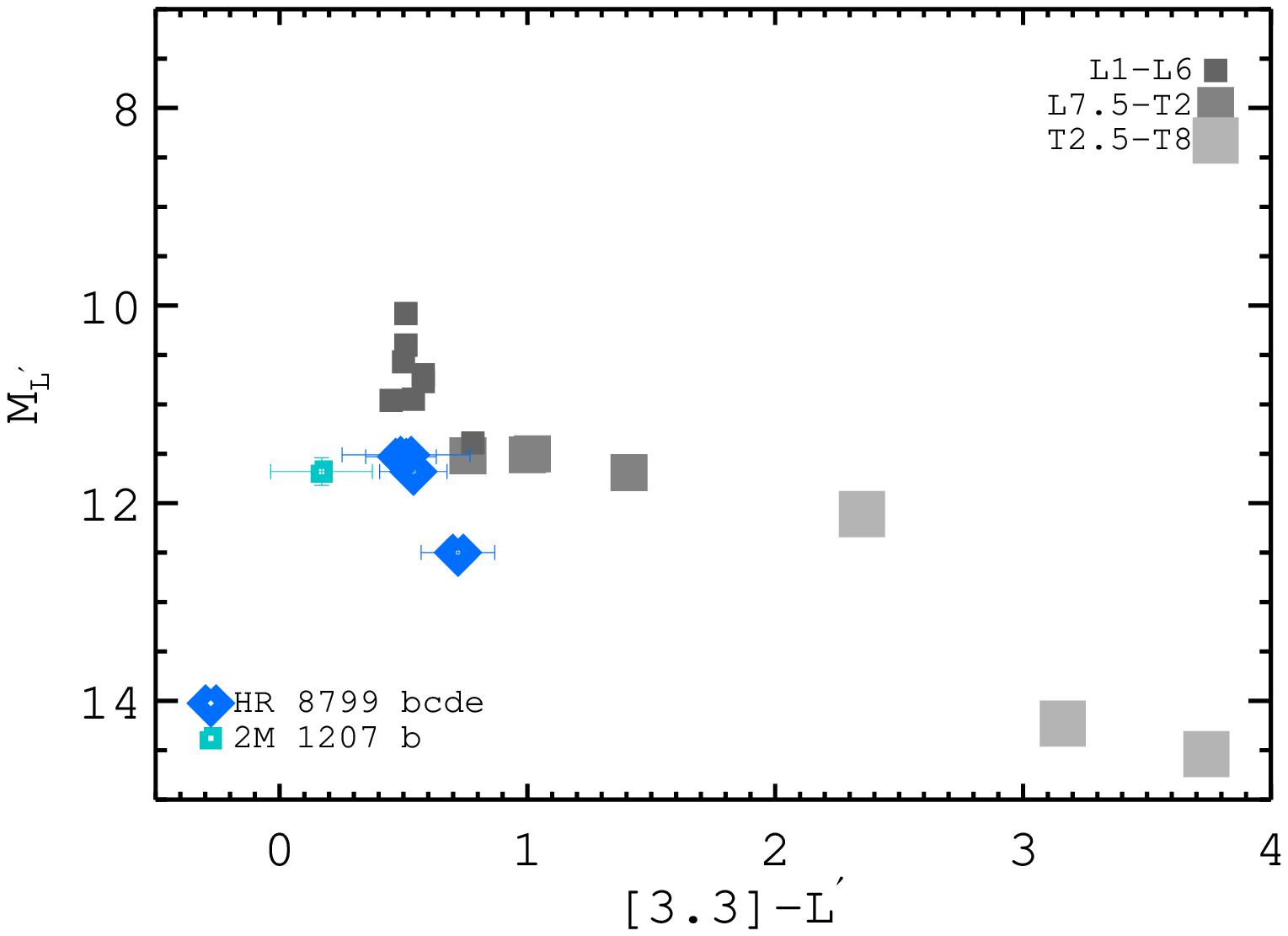}
\includegraphics[scale=0.52,trim=10mm 5mm 5mm 0mm,clip]{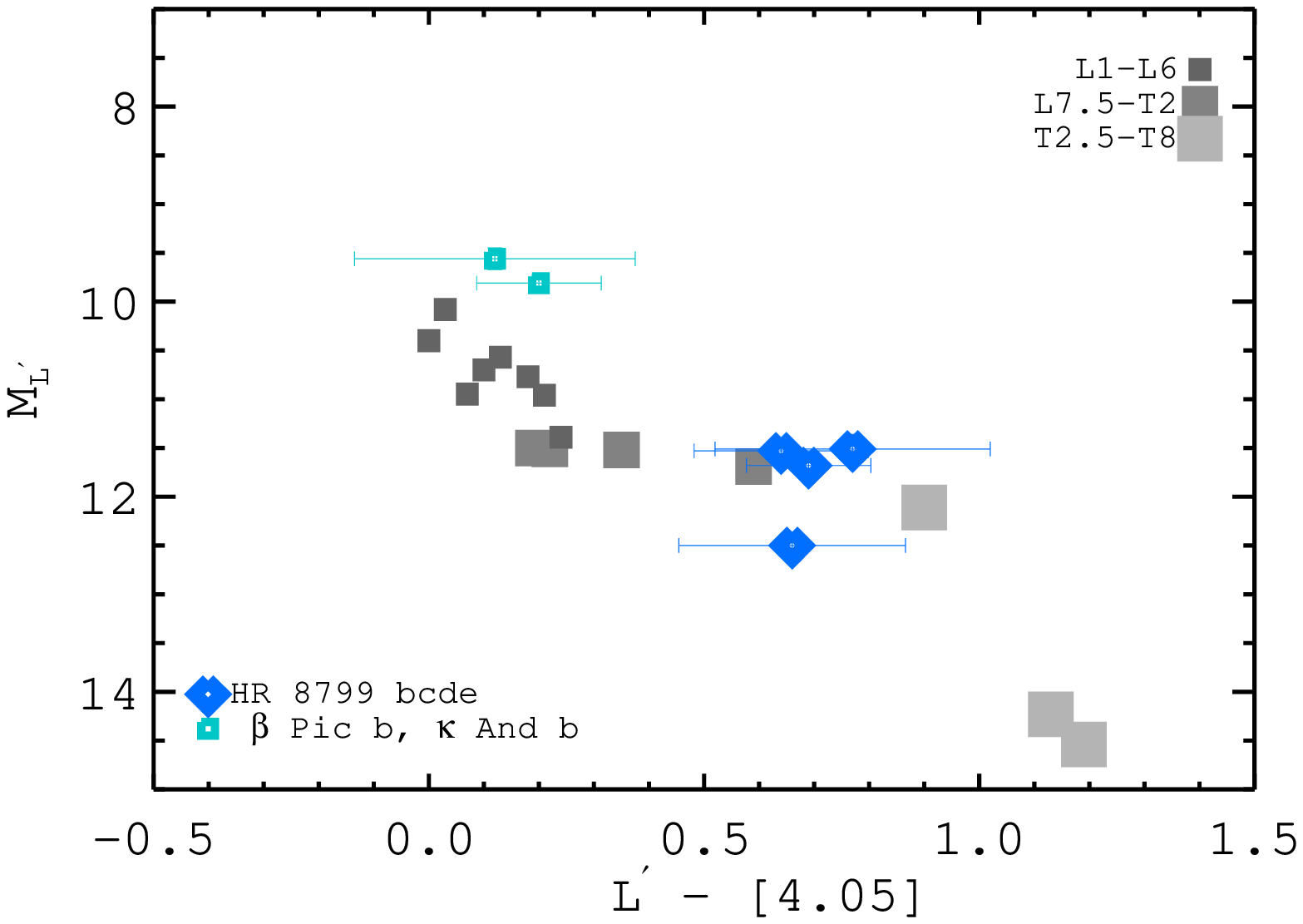}
\caption{Color-magnitude diagrams comparing HR 8799 bcde to field brown dwarfs and other planet-mass companions in $K_{s}$ vs $H$-$K_{s}$ (top-left), $L^\prime$ vs. $H$-$L^\prime$ (top-right), $L^\prime$ vs. [3.3] - $L^\prime$ (bottom-left), and $L^\prime$ vs. $L^\prime$ - [ 4.05] (bottom-right).  For the \citeauthor{Leggett2010} sample of field objects we use circles; for the (smaller, mid-IR focused) \citeauthor{Sorahana2012} sample we use squares.}
\label{color-color}
\end{figure}

 Figure \ref{color-color} compares HR 8799 bcde's near-to-mid infrared colors to those for field brown dwarfs and other imaged planets/planet-mass companions.   The $K_{s}$ vs. $H$-$K_{s}$ (top-left) and $L^\prime$ vs. $H$-$L^\prime$ diagrams (top-right) show that the near-IR colors of young, directly-imaged planetary companions appear to lie off of the field dwarf locus: hotter ($T_{\rm{eff}}$ $\sim$ 1600--2000 $K$) companions like $\beta$ Pic b and ROXs 42Bb appear redder than the dwarf sequence, while cooler ($T_{\rm{eff}}$ $\sim$ 800-1200 $K$) companions HR 8799 bcde, HD 95086 b, and 2M 1207 B appear to define a reddened extension of the L dwarf sequence to fainter magnitudes more typical of T dwarfs \citep[e.g.][]{Marois2008,Currie2011a,Barman2011,Galicher2014,Skemer2014}.   Likewise, HR 8799 bcde and 2M 1207 B appear to follow an extension of L dwarf-like [3.3]-$L^\prime$ colors to fainter $L^\prime$ magnitudes (bottom-left), exhibiting rather flat 3--4 $\mu m$ spectra consistent with weaker methane absorption \citep{Skemer2012,Skemer2014}.   
 
 In contrast, the $L^\prime$ - [4.05] colors for HR 8799 bcde (bottom-right) may appear more ``'normal", consistent with the field sequence.  While the spectral type sampling for AKARI-selected brown dwarfs is sparser than for the \citeauthor{Leggett2010} sample providing comparisons at shorter wavelengths,  HR 8799 cde appear to lie in a region consistent with early field T dwarfs.  HR 8799 b's colors are harder to parse, although its position is consistent (within errors) to slightly later T dwarfs.

\subsection{Color Variations in the HR 8799 planets}

Including our new [4.05] photometry, HR 8799 bcde now have high-quality photometric measurements in five passbands, from 1.6 $\mu m$ to $\sim$ 4 $\mu m$, HR 8799 bcd have detections at $J$ and $M^\prime$, while HR 8799 b is marginally detected at $z/Y$ band \citep{Marois2008,Galicher2011,Currie2011a}.    To determine
 whether the planets' photometry and colors differ at a significant level, we follow methods performed in \citet{Currie2013}, quantifying the agreement between their photometry using the reduced $\chi^{2}$ and goodness-of-fit statistics, considering photometric errors for both the given planet and its comparison, and adopting the zero-point flux densities listed in \citet{Currie2013}.    
 To assess instead whether or not the planets are systematically fainter/brighter than one another but have identical colors, we then compare scaled versions of the planets' photometry, adopting the scaling factor that minimizes the $\chi^{2}$ value between the pairs of planets.
 
 For comparisons between HR 8799 bcd, we consider measurements at $J$,$H$,$K_{s}$,$[3.3]$,$L^\prime$,$[4.05]$, and $M^\prime$.   Comparisons including HR 8799 e focus on $H$,$K_{s}$,[3.3],$L^\prime$, and $[4.05]$.   To assess how our results depend on measurements from different groups, we separately consider $H$ band HR 8799 bcd photometry from \citet{Currie2012b} and $JH$ HR 8799 bcde photometry from \citet{Oppenheimer2013}  \footnote{We do not consider $K_{s}$ band photometry from \citet{Esposito2012} since the photometric errors reported are equal to or much larger than errors reported by \citet{Marois2008,Marois2010a} and \citet{Currie2011a}.  Similarly, we do not consider [3.3] photometry for HR 8799 bc from \citet{Currie2011a}, which is 0.4-0.7 mags fainter than reported in higher signal-to-noise data in \citet{Skemer2012}.   The detections of both planets in the \citet{Currie2011a} reduction are secure and statistically significant:  \citet{Currie2011a} report a detection of HR 8799 b at SNR $\sim$ 3.8, while our revised value, using methods employed in this paper and identical to those widely adopted by other authors \citep[e.g.][]{Marois2008,Lafreniere2007b} yields a slightly higher, SNR $\sim$ 5.  Thus, the significance of the detection is not "overly optimistic" as suggested in \citet{Skemer2012}.  However, the photometric calibration was hampered by clouds affecting the unsaturated star observations we used for flux calibration and the signal loss due to processing was slightly underestimated.  Since the \citet{Skemer2012} is better calibrated and verified (in part) from follow-up work \citep{Skemer2014} we simply adopt it here.}.

 Table \ref{tab_var} presents  our results.  For our nominal photometry, HR 8799 b and e exhibit clear differences (C.L. $>$ 0.98) in (scaled) photometry.    In contrast, we can only demonstrate that HR 8799 d and e have a higherluminosity thanHR 8799 b, not that they have different colors, probably because of the higher $J$ band photometric error/non-detection for HR 8799 d and e.   The best-fit scaling factor applied to HR 8799 c when compared to d and e is always greater than unity, consistent with it being a somewhat lower-luminosity planet (scaling factor $\sim$  0.82-0.92 or $\sim$ 10-20\% fainter).  However, we do not find a statistically significant difference in the photometry or colors between HR 8799 cde, with or without flux scaling.  
 
Adopting the \citet{Currie2012b} $H$ band photometry does not change our results: HR 8799 b's photometry and colors differ from c's, b's photometry and colors differs from de's, while we cannot identify significant differences between HR 8799 cde.   In contrast, the \citet{Oppenheimer2013} P1640 $J$ and $H$ photometry suggest that HR 8799 b and c's colors do not differ at a greater than 2 $\sigma$ level.
  This disagreement is in part due to the substantially fainter P1640 $J$ band photometry for HR 8799 c, which leaves the planet with $J$-$H$ colors more similar to HR 8799 b than implied from other sources \citep{Marois2008,Currie2012b,Skemer2012}.    Adopting the P1640 $J$ and $H$ measurements does not reveal any clear differences in colors between HR 8799 cde.  The far brighter $H$-band photometry for HR 8799 e compared to c do appear to diverge at the 92\% confidence limit, although this difference is less than 2-$\sigma$ and thus not considered here to be significant.

\subsection{Model Atmosphere Comparisons to HR 8799 \lowercase{bcde}}

Combining our [4.05] photometry, $L^\prime$ photometry and other mid-IR photometry and comparing them to models provides some new insights into HR 8799 bcde's atmospheric properties.  

\begin{figure}
\centering
\includegraphics[scale=0.35]{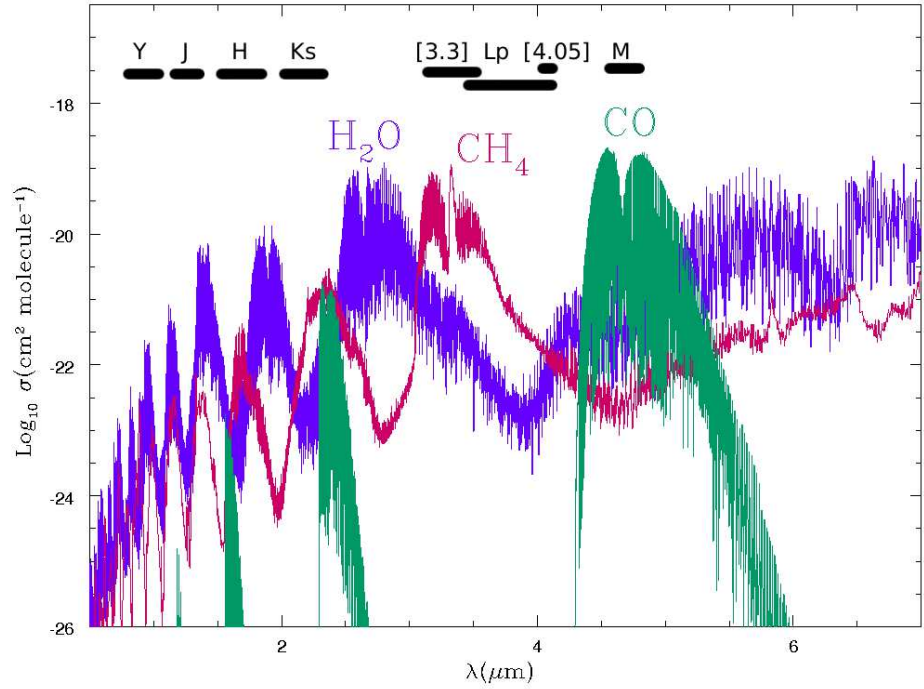}
\caption{Absorption cross section per molecule for major molecular opacity sources -- $H_{2}O$, $CH_{4}$, and $CO$ -- calculated from \citet{SharpBurrows2007} at conditions relevant for young self-luminous super-jovian planets: 1200 $K$ and 1 bar atmospheric pressure.  Trends covering the full range of $T_{eff}$ for young planets HR 8799 bcde and HD 95086 b (900--1400 $K$) are similar.  We over plot the major near-to-mid infrared passbands.}
\label{cross-section}
\end{figure}

The [4.05] filter in particular serves as a useful point-of-reference to clarify how chemistry and clouds shape a young planet's atmosphere across mid-infrared wavelengths (3--5 $\mu m$).  
Figure \ref{cross-section} compares the sensitivity of young planet photometry in different passbands to major molecular opacity sources -- $H_{2}O$, $CH_{4}$, and $CO$.   $CH_{4}$ and $CO$ have high absorbing cross-sections at the [3.3] and $M$ passbands, respectively, and thus photometry in these filters is strongly affected by non-equilibrium carbon chemistry \citep[e.g.][]{Galicher2011,Skemer2012,Skemer2014}.  Additionally, $CH_{4}$ is strong at the blue end of the $L^\prime$ passband, rendering $L^\prime$ photometry sensitive to carbon chemistry as well \citep[c.f. Figure 6a in][]{Skemer2012}.  In contrast, the [4.05] filter resides in a region of low opacity for major species $CH_{4}$, $H_{2}O$, and $CO$ as well as $CO_{2}$ \citep{Sorahana2012}.   The [3.3] to [4.05] SED and $L^\prime$-[4.05] color vary with different levels of methane depletion: e.g. heavier depletion results in a flatter $L^\prime$-[4.05] slope. 

Additionally,  the $L^\prime$-[4.05] flux varies with cloud thickness.  Thicker clouds make this slope slightly bluer (c.f. Figures 4, 7, and 10 in Madhusudhan et al. 2011).   ``Patchy" cloud regions may yet make the $L^\prime$-[4.05] slope redder again \citep[see][]{Currie2011a}.    As we will show, the [4.05] photometry may strengthen the case for thicker, patchy clouds  considered in \citet{Currie2011a} (this section), while disfavoring other, recent best-fitting models (section  6.2).

\subsubsection{Modeling Goals and Formalism}
HR 8799 bcde's photometry and/or spectra are notoriously difficult to model.  Those few models successfully reproducing at least most of the available data and implying physically realistic radii/masses require thick clouds and non-equilibrium carbon chemistry \citep{Currie2011a, Madhusudhan2011, Galicher2011, Skemer2012}.   Additionally, the best-fit models thus far require the clouds to be patchy \citep{Currie2011a,Skemer2014}, although the fits are not perfect and heterogeneous cloud models are still in their early stages of development.    The fit quality may also depend on other key assumptions, including dust grain sizes and size distributions \citep[e.g.][]{Currie2013,Currie2014b} and complete molecular line lists for important species \citep{Yurchenko2014}, which were absent in the current best-fitting models.   New, soon-to-be acquired direct spectroscopy for HR 8799 cde may also identify new discrepancies with a model best fitting the current suite of data.  

Therefore, we do not perform a rigorous parameter search to quantitatively identify the best-fitting models as in previous work \citep{Currie2011a,Madhusudhan2011,Currie2013,Currie2014b}.  Rather, we compare new data to recent best-fitting models from A. Burrows \citep{Currie2011a,Madhusudhan2011,Skemer2012}, then an expanded set of the thick cloud, non-equilibrium chemistry Burrows models considered in \citet{Skemer2012}, and finally a set of these models approximating patchy, thick clouds  to identify new challenges to reproducing HR 8799 bcde's photometry/spectra and deriving its physical properties.    Given the difficulties in constructing self-consistent patchy cloud models, we simply follow \citet{Currie2011a} and \textit{approximate} a patchy atmosphere as a linear combination of two model atmospheres with the same $T_{\rm{eff}}$, surface gravity, and radius but different cloud scale heights\footnote{Note that this differs from the method used in \citet{Skemer2012,Skemer2014}.  There, they use a linear combination of atmospheres with \textit{different} $T_{\rm{eff}}$ in addition to different cloud scale heights.   As model atmospheres with thicker clouds have hotter \textit{P}-\textit{T} profiles (i.e. the atmosphere is hotter at a given pressure), this approach may result in a more uniform \textit{P}-\textit{T} profile across the planet surface, depending on exactly which atmosphere models are used.  However, the Burrows models used are explicitly tied to radii from the \citet{Burrows1997} planet evolutionary models: e.g. the 700 $K$ and 1400 $K$ model atmospheres comprising \citeauthor{Skemer2012}'s patchy atmosphere approximation differ in radius (by just under $\sim$ 10\%) and are therefore discontinuous.  In our approach, the radii for both atmosphere components are the same and the \textit{P}-\textit{T} is implicitly discontinuous across the boundary regions separating the components.}.  

Table \ref{modelused} lists the model atmospheres used for each SED comparison.    Briefly, the A-type cloud models are the thickest, significantly more vertically extended than models (ÒE-typeÓ) used to compare with field brown dwarfs \citep{Burrows2006}, resulting in a hotter pressure-temperature profile.  The AE-type clouds are just slightly thinner than those in the A-type models: while the A-type clouds track the gas density profile as a whole, the AE-type clouds' scale heights are reduced by a factor of two.  The models consider modal dust grain sizes of 30 and 60 microns (e.g. ``A60" refers to a model with the ``A" cloud type and modal dust grain sizes of 60 $\mu m$).  Lastly, the models consider both atmospheres in chemical equilibrium and those in disequilibrium, where we modify the abundance ratios of CO to CH4 .  For example, 100CO, 0.01 CH$_{4}$ denotes a model with abundances of CO (CH$_{4}$) 100 times more (less) than solar.  See previous studies for more details\citep{Burrows2006,Madhusudhan2011,Currie2011a,Currie2013,Currie2014b}.

For simplicity and because we fail to identify clear differences amongst HR 8799 cde's photometry, we display comparisons only for HR 8799 b and d for most models.  We include comparisons for all four HR 8799 planets only after identifying the modeling approach that appears to best reproduce the outer two planets' data.   We consider planet radii between 0.9 and 1.15 times the nominal radius assumed in each model ($R$ $\sim$ 1.1--1.5 $R_{J}$).  

\subsubsection{Modeling Results}
\begin{figure}
\centering
\includegraphics[scale=0.55,trim=10mm 2mm 6mm 0mm,clip]{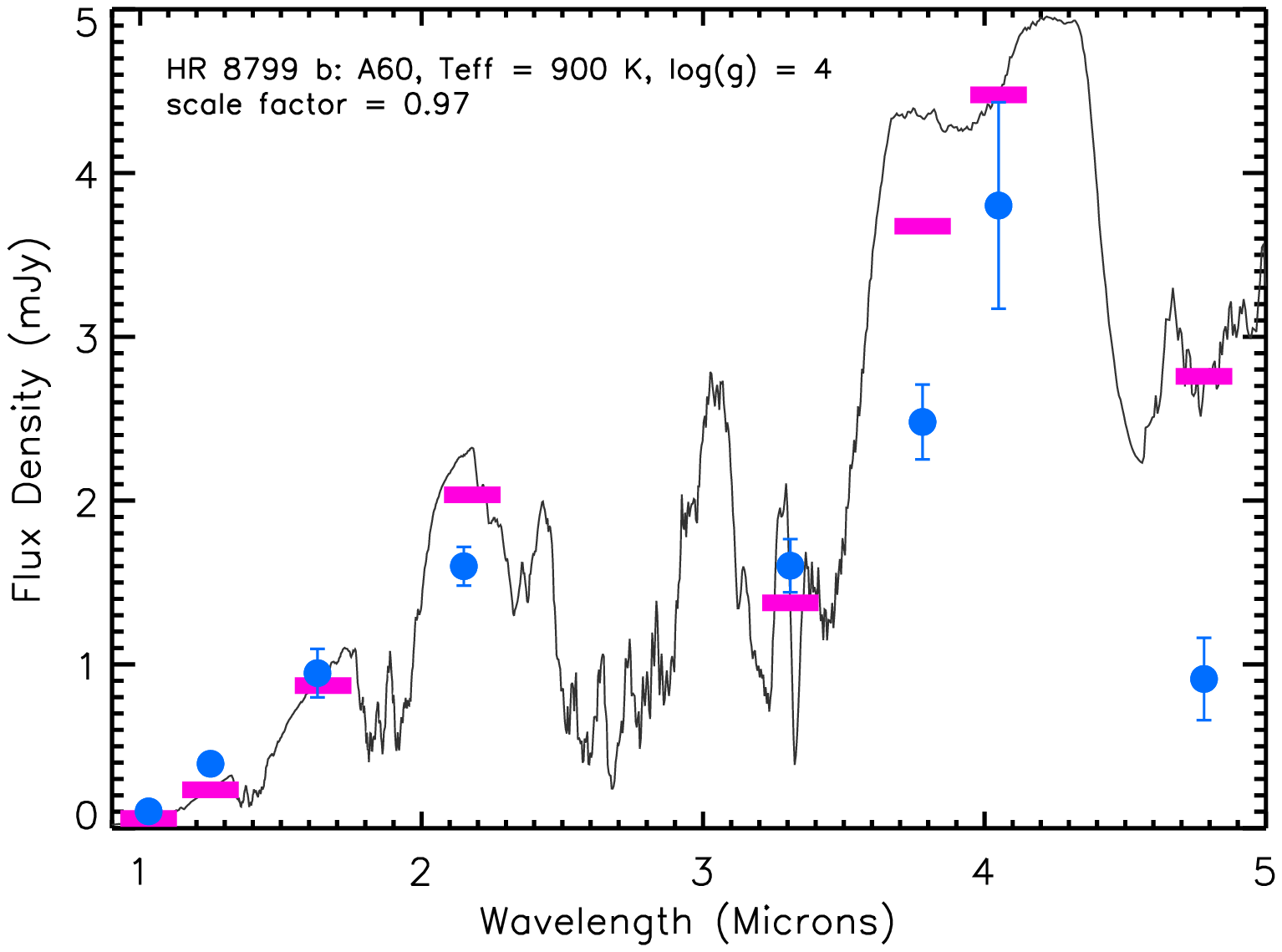}
\includegraphics[scale=0.55,trim=22mm 2mm 5mm 0mm,clip]{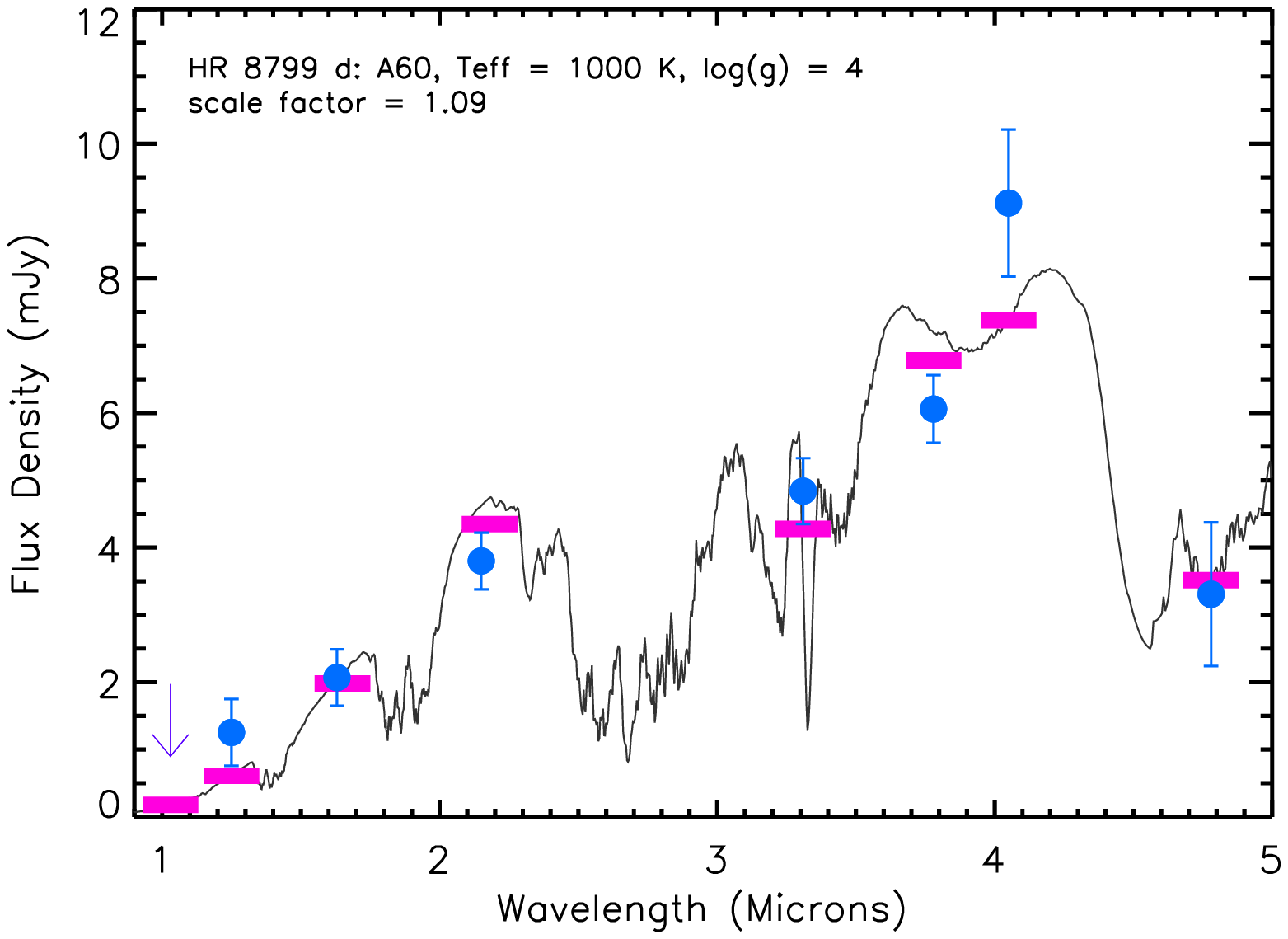}
\\
\includegraphics[scale=0.55,trim=10mm 2mm 6mm 0mm,clip]{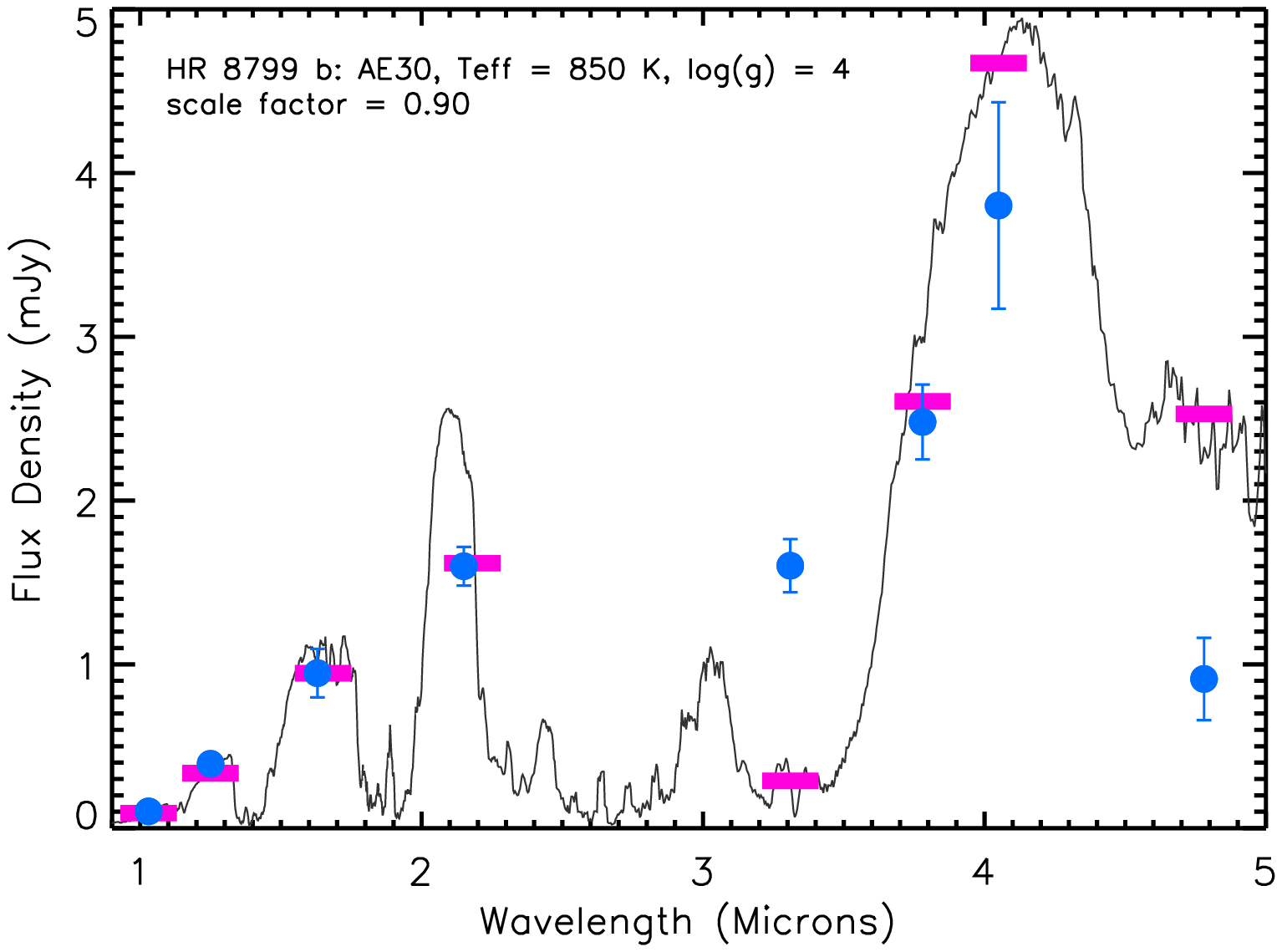}
\includegraphics[scale=0.55,trim=22mm 2mm 5mm 0mm,clip]{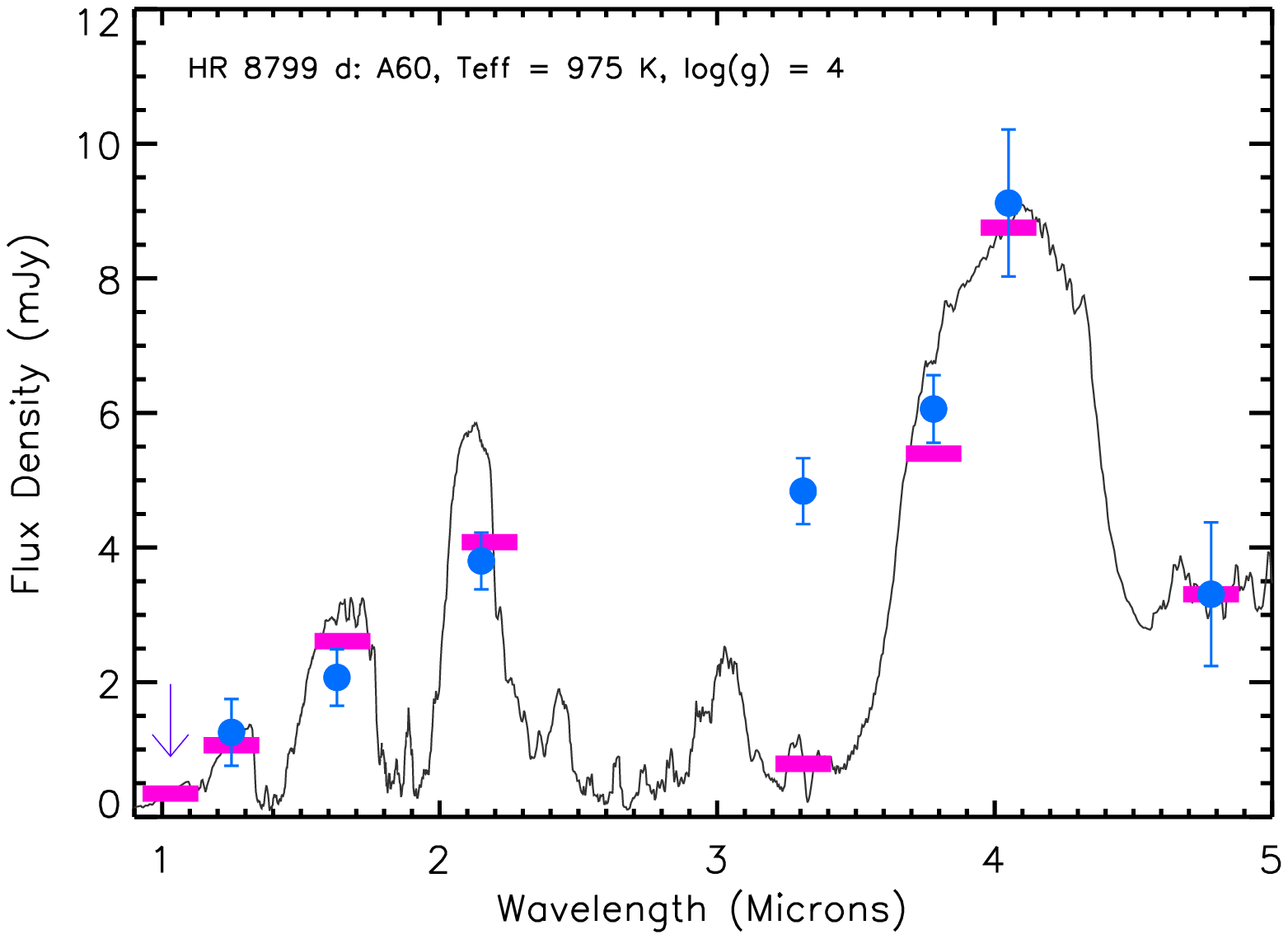}
\\
\includegraphics[scale=0.55,trim=10mm 2mm 6mm 0mm,clip]{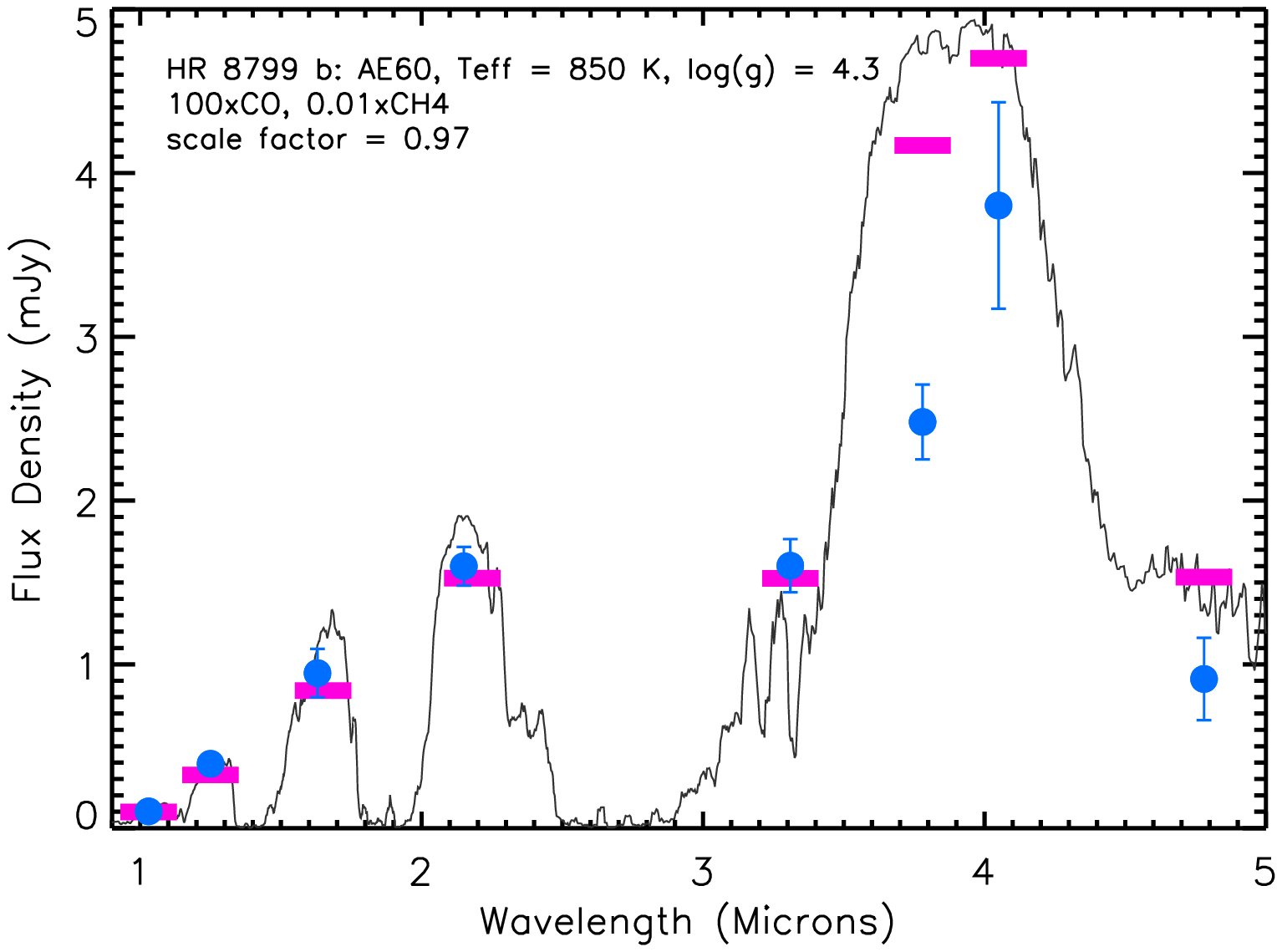}
\includegraphics[scale=0.55,trim=22mm 2mm 5mm 0mm,clip]{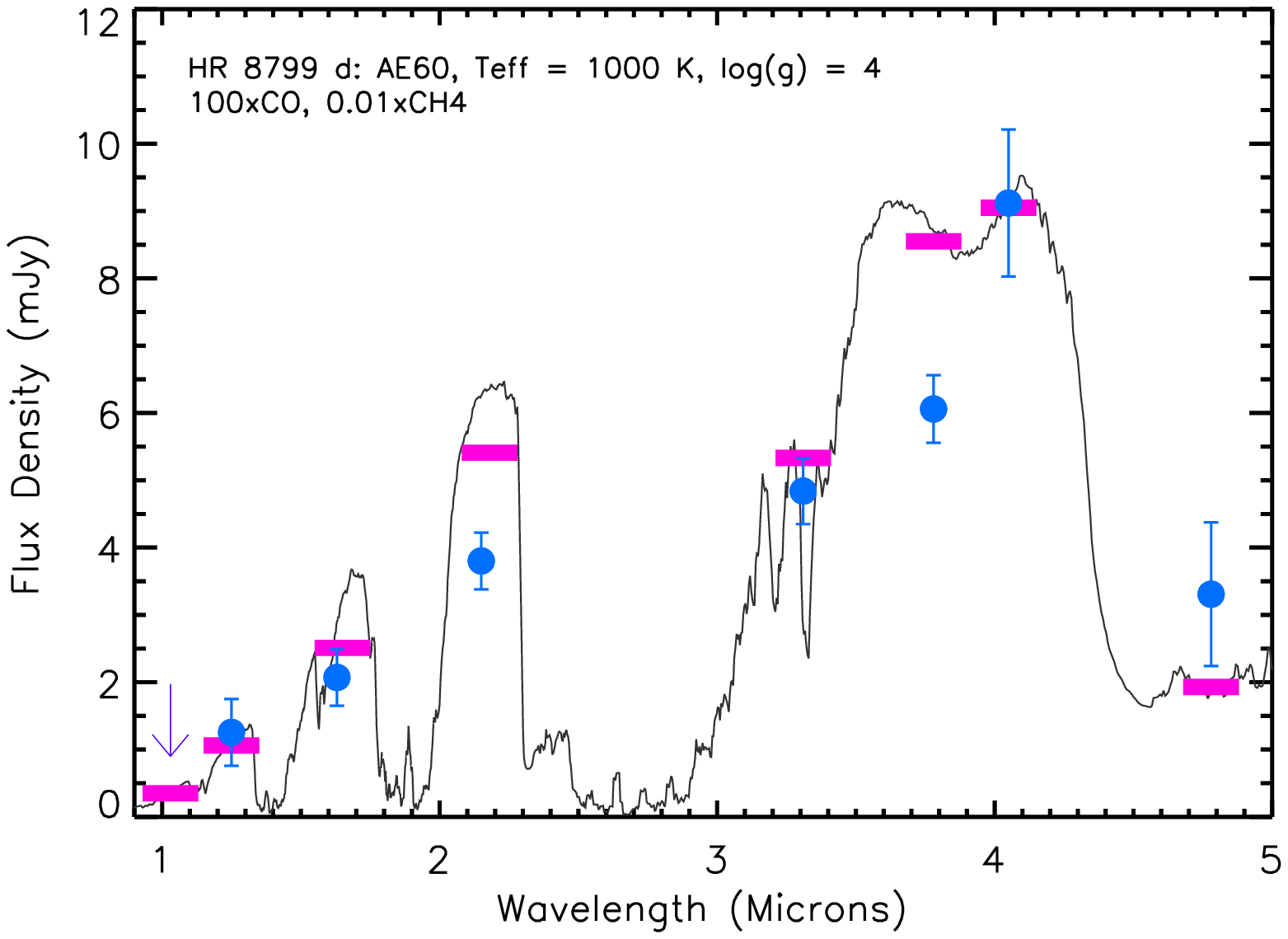}
\caption{Atmosphere model comparisons for HR 8799 b (lefthand panels) and HR 8799 d (righthand panels) using the thick cloud, chemical equilibrium models from \citet{Currie2011a} (A-type clouds, modal particle size of 60 $\mu m$; top panels); the thick cloud, chemical equilibrium models from \citet{Madhusudhan2011} (AE-type clouds, modal particle size of 60 $\mu m$; middle panels); and the thick cloud, non-equilibrium carbon chemistry models from \citet{Skemer2012} (also with AE-type clouds, modal particle size of 60 $\mu m$).  Blue dots represent detections; downward pointing arrows represent 5-$\sigma$ upper limits.  Horizontal magenta lines represent the filter-convolved flux densities.  The models shown are those, based on visual inspection, that best match the HR 8799 bd photometry.  The ``scale factor" refers to the constant by which we multiply each atmosphere model's radius: for panels without this entry, the scale factor is 1 (i.e. no rescaling of the assumed model atmosphere radius). The specific models used are labeled in the figure captions and listed in Table \ref{modelused}: see \citet{Burrows2006}, \citet{Currie2011a}, \citet{Madhusudhan2011}, and \citet{Skemer2012} for more details on the model input physics and terminology.}
\label{seds1}
\end{figure}

Figure \ref{seds1} compares HR 8799 bd to the thick cloud, chemical equilibrium models in \citet{Currie2011a} and \citet{Madhusudhan2011} and the thick cloud, non-equilibrium chemistry models in \citet{Skemer2012}.   All three models successfully reproduce the planets' weak/suppressed emission at 1--2 $\mu m$.   The  chemical equilibrium model from \citet{Madhusudhan2011} successfully reproduces the planets' $L^\prime$ photometry and (nearly) matches the peaks of the spectra at 4.05 $\mu m$ but significantly underpredicts the 3.3 $\mu m$ photometry for both planets while overpredicting HR 8799 b's $M^\prime$ flux density.   The \citet{Currie2011a} models invoking even thicker clouds overpredicts the HR 8799 b $L^\prime$ measurement, matches both planets'  3.3 $\mu m$ photometry, but again overpredicts HR 8799 b's $M^\prime$ photometry.  In contrast, the non-equilibrium chemistry models overpredict both planets'  $L^\prime$ photometry and HR 8799 b's [4.05] photometry.  

Thick but patchy cloud approximations (Figure \ref{seds2}) may significantly improve fits to all four HR 8799 planets.   While the AE60 chemical equilibrium and non-equilibrium models depicted in Figure \ref{seds1} and used in \citet{Madhusudhan2011} and \citet{Skemer2012} fail to reproduce multiple HR 8799 b photometric points between 3 and 5 $\mu m$, our patchy cloud approximation overpredicts only the $M^\prime$ photometry.    Despite the apparent improved fit using this chemical equilibrium, patchy cloud approximation, we fail to identify any additional model variable -- modal dust grain size, cloud scale height, metallicity, surface gravity, etc. -- \textit{except} for non-equilibrium carbon chemistry/non-solar $CO$/$CH_{4}$ ratios that could drive down the predicted $M^\prime$ flux density.  

In contrast, the models first matched to HR 8799 data in \citet{Currie2011a} fare significantly better in reproducing HR 8799 cde's photometry.   The A60 thick cloud, chemical equilibrium model appears to reproduce HR 8799 d photometry better than either AE60 model (Figure \ref{seds1}, right panels).  HR 8799 ce  also appear to be better fit by the A60 thick cloud models, expected since HR 8799 ce have broadband colors indistinguishable from HR 8799 d.    Furthermore, agreement with the data for all three inner planets may improve if 15--35\% of HR 8799 cde's atmosphere is covered by thinner clouds (Figure \ref{seds2}, top-right panel and bottom panels).   Non-equilibrium carbon chemistry then does not appear to be strictly required to explain the inner three planets' photometry.  While HR 8799 c's $K$-band spectrum shows clear evidence for methane depletion and enhanced carbon monoxide \citep{Konopacky2013}, the d and e planets lack published $K$-band spectra.
\section{Discussion}
\subsection{Detecting a Fifth Planet Orbiting Interior to HR 8799 e}
Our study presents the first detections of HR 8799 bcde in the [4.05] narrowband filter and deep, multi-epoch detections of the four planets at $L^\prime$.   Although our Keck/NIRC2 $L^\prime$ data yield (some of) the most sensitive imaging limits on planets at small, 1.5--4 $\lambda$/D separations ($\sim$ 0\farcs{}14--0\farcs{}37 or $\sim$ 5-15 $AU$), we fail to detect a fifth planet.  Our limits are inconsistent with an additional HR 8799 cde-luminosity/mass object at r$_{proj}$ $\sim$ 9--15 $AU$, or within/exterior to the predicted location of HR 8799's warm dust belt, and an HR 8799 b-like object exterior to 12 $AU$.   At smaller separations, planets like HR 8799 bcde are undetectable, but our data rule out a more massive (12--13 $M_{J}$) companion at the inner edge of the warm dust belt ($r$ $\sim$ 5--6 $AU$).   

Our detection limits for an HR 8799 f complement those derived by \citet{Hinkley2011} using sparse aperture masking also at $L^\prime$.   Those data are uniquely sensitive to planets at 3-5 AU, interior to our inner working angle.   At 5-6 AU our contrasts and thus mass limits are comparable, while ours are deeper at wider separations, especially exterior to 0\farcs{}22.  Together, they set stringent limits on the presence of planets more massive than HR 8799 bcde located interior to HR 8799 e.

Our limits are broadly consistent with the phase space \citet{Gozdziewski2014} find possible for an ``HR 8799 f" in a 3:1 or 2:1 mean-motion resonance with HR 8799 e.  We fail to find an additional planet at their predicted positions of 7.5 $AU$ and 9.7 $AU$ ($r_{proj}$ $\sim$ 7--9 $AU$ $\sim$ 0\farcs{}17--0\farcs{}22).  However, in their model, ``HR 8799 f" likely has a mass of 2--8 $M_{J}$ at 7.5 $AU$  and 1.5--5 $M_{J}$ at 9.7 $AU$, which would not be detectable at the 5-$\sigma$ level even in our most sensitive data set \footnote{Still, the best-predicated locations and masses  for a 5th planet should be updated as our knowledge of the orbital properties of HR 8799 bcde (and thus the stability of the entire system) grows.  For instance, 
\citet{Gozdziewski2014} model HR 8799 bcde as having coplanar orbits inclined by $\approx$ 25-28$^{o}$ to the sky.   In contrast, modeling astrometry for HR 8799 bcd between 1998 and 2010 instead may favor more highly inclined orbits at least for HR 8799 d and, possibly, slightly \textit{non}-coplanar orbits \citep{Currie2012b}.}.  

Extreme adaptive optics, high-contrast imaging well beyond what is capable with instruments like NIRC2 is required to image ``HR 8799 f" at its predicted location and full mass range.   For example, the $L^\prime$ contrast required to image a 2 $M_{J}$ planet at $r_{proj}$ $\sim$ 7 $AU$ around HR 8799 is $\sim$ 1$\times$10$^{-5}$ \citep[e.g. from the BT-Settl models][]{Allard2012}, or about 80 times fainter than our 5-$\sigma$ limit with NIRC2 $L^\prime$ or the nominal LBTI/LMIRCam performance \citep{Skemer2014b}.  The challenge at $H$ is even steeper, requiring a contrast of 5$\times$10$^{-7}$, or about 800 times deeper than the best $H$-band contrasts at $r$ $<$ 0\farcs{}3 with a conventional AO system \citep[e.g.][]{Brandt2014}.  It is also about 200 times deeper than first-light \textit{Gemini Planet Imager} (GPI) contrasts at $r$ $\sim$ 0\farcs{}2 on southern stars \citep{Macintosh2014}.  The \textit{Subaru Coronagraphic Extreme Adaptive Optics Imager} (SCExAO) \citep{Martinache2009} may be the only upcoming high-contrast imaging platform capable of searching for a fifth jovian planet orbiting HR 8799 with a mass just above that of Jupiter, as it is expected to yield very high contrasts within several diffraction beam widths of the star ($r$ $\sim$ 0\farcs{}1--0\farcs{}2 at $H$-band ).

\subsection{Constraining HR 8799 bcde Atmospheric Properties and Chemical Diversity}
Our data also yield the first photometric detections of HR 8799 bcde in the Br-$\alpha$/[4.05] filter.
While HR 8799 bcde appear ``red/underluminous" in near-infrared colors compared to field L/T type brown dwarfs of similar temperatures \citep[800--1200 $K$][]{Marois2008,Currie2011a} and blue/luminous at 3.3 $\mu m$  \citep{Skemer2012}, the planets' $L^\prime$-[4.05] colors might not depart from the field L/T dwarf sequence.   Within the field dwarf context, their $L^\prime$-[4.05] colors may appear early/mid T dwarf-like in contrast to the more L dwarf like near-IR properties.

\begin{figure}
\centering
\includegraphics[scale=0.55,trim=10mm 2mm 6mm 0mm,clip]{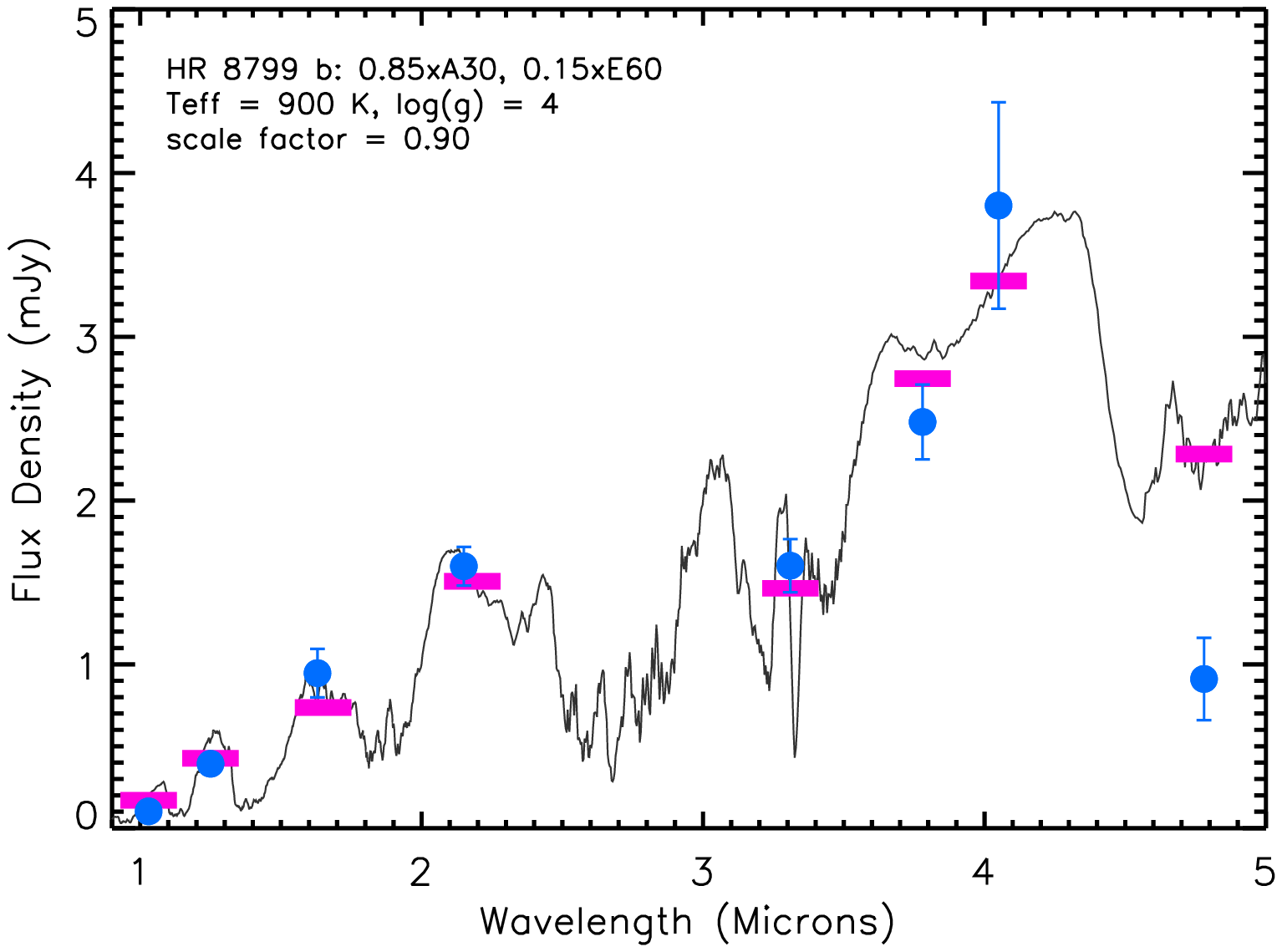}
\includegraphics[scale=0.55,trim=22mm 2mm 5mm 0mm,clip]{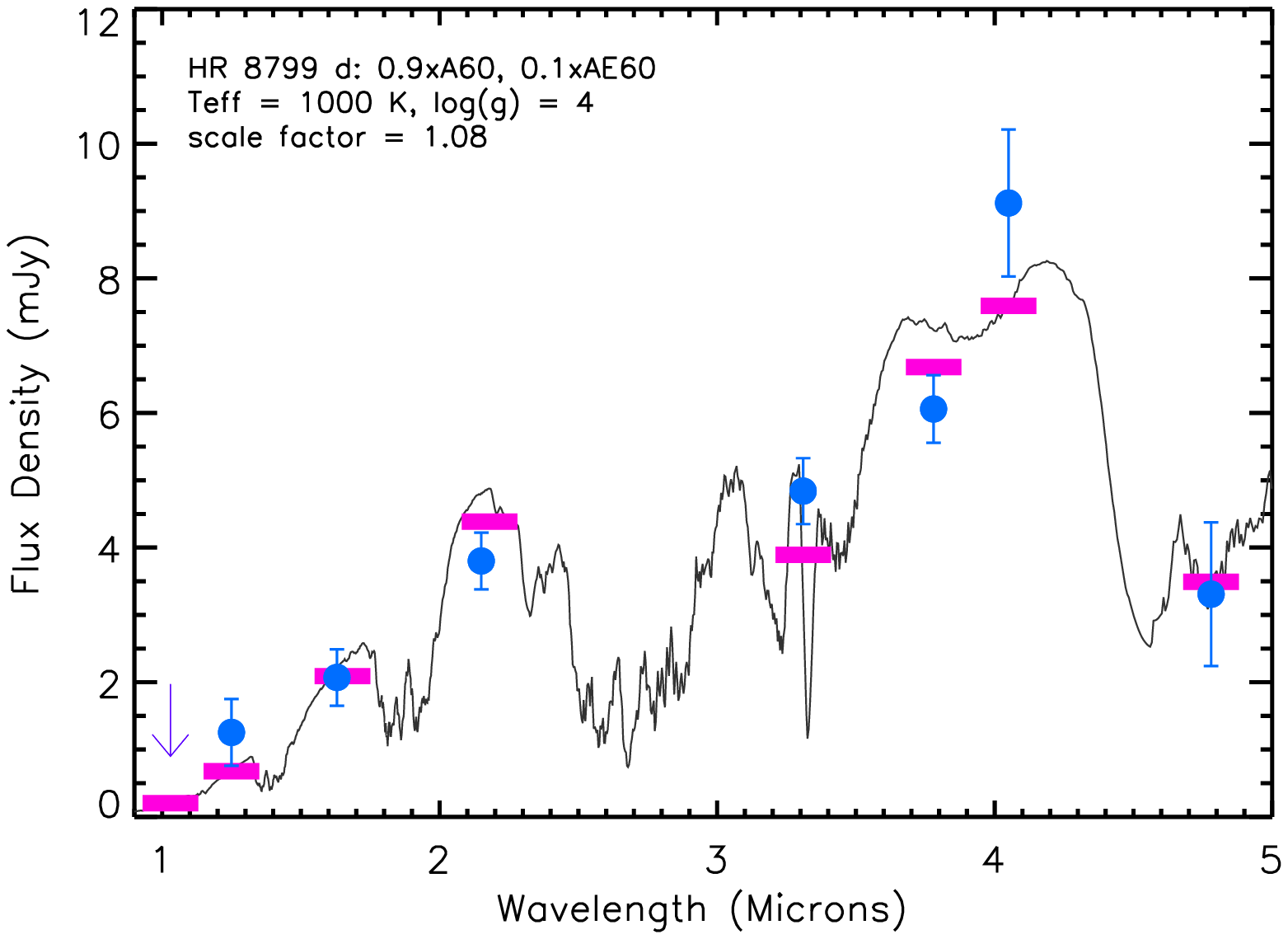}
\\
\includegraphics[scale=0.55,trim=10mm 2mm 6mm 0mm,clip]{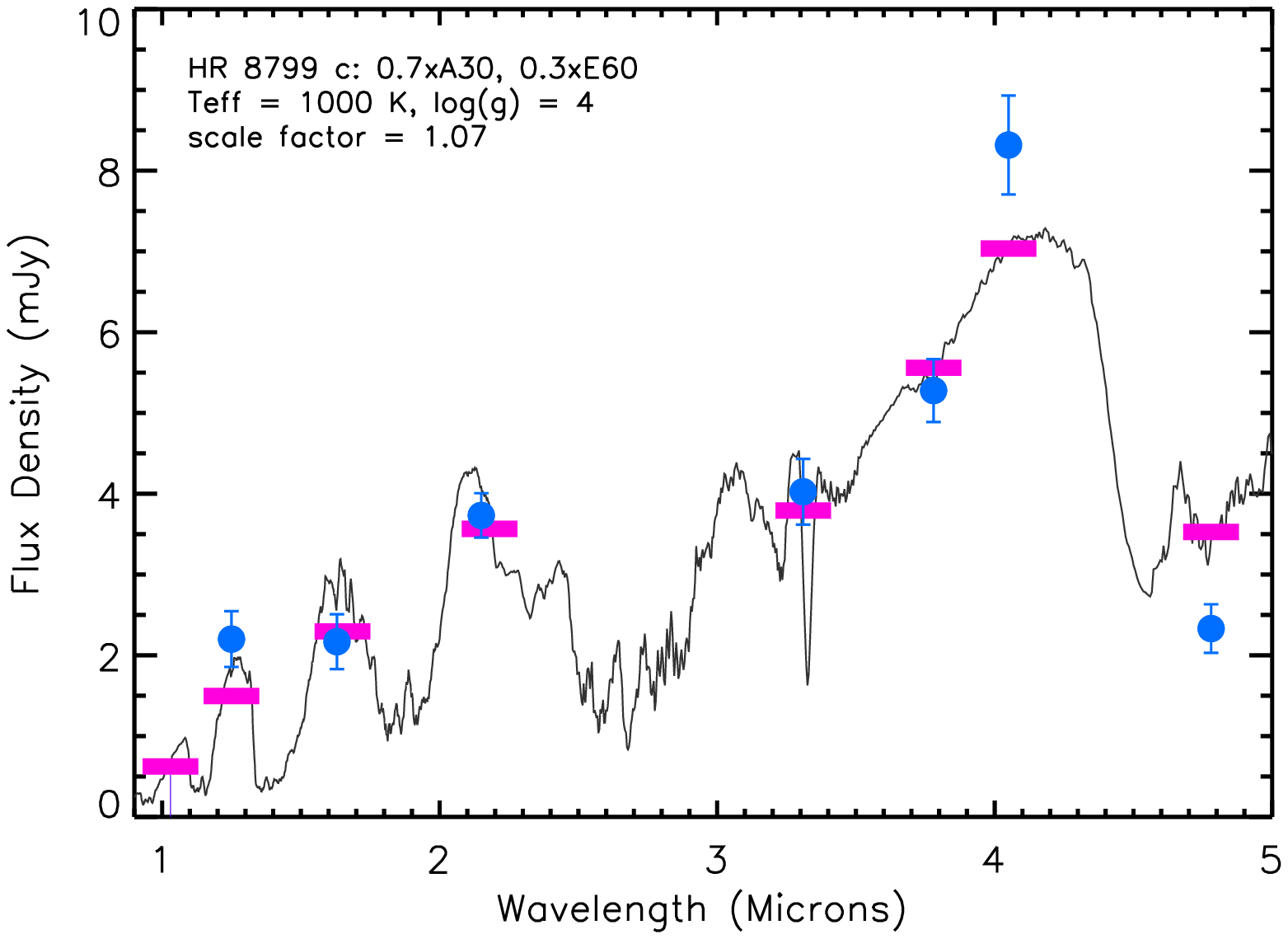}
\includegraphics[scale=0.55,trim=22mm 2mm 5mm 0mm,clip]{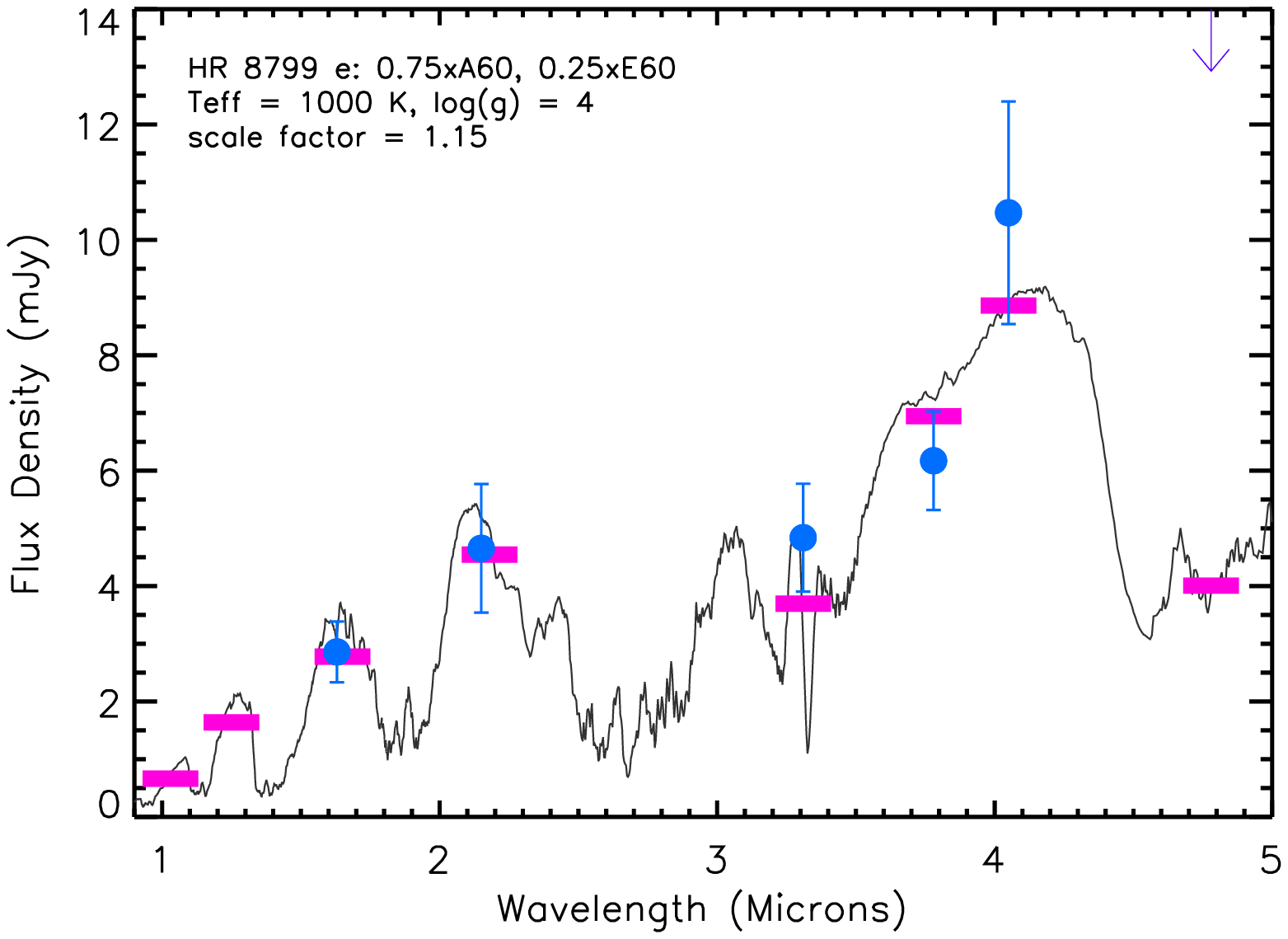}
\caption{Atmosphere model comparisons for HR 8799 b (top-left panel), HR 8799 c (bottom-left panel), HR 8799 d (top-right panel) and HR 8799 e (bottom-right panel) using the thick, patchy cloud, chemical equilibrium approximation.  Symbols are the same as in Figure \ref{seds1}.  The relative fractions of the thicker to thinner cloud components for each model atmosphere approximation (i.e. relative fraction of A60 to AE60 or E60 components) are as follows:  0.85, 0.15 (HR 8799 b); 0.7, 0.3 (HR 8799 c);  0.9, 0.1 (HR 8799 d); and 0.75, 0.25 (HR 8799 e).}
\label{seds2}
\end{figure}

Our new [4.05] photometry complicates atmospheric modeling efforts.  HR 8799 bcde's very red $L^\prime$-[4.05] color/rising SED from 3.8 to 4 $\mu m$ is poorly reproduced by non-equilibrium chemistry models invoked by \citet{Skemer2012} to explain the planets' 3.3 $\mu m$ and $M^\prime$ photometry.   Conversely, the AE60 thick cloud models presented in \citet{Madhusudhan2011} accurately reproduce HR 8799 bcde's $L^\prime$ - [4.05] color but are poor matches at 3.3 $\mu m$ and $M^\prime$.   The suite of thick (patchy) cloud atmosphere models from \citet{Marley2010} considered by \citet{Skemer2014} to match HR 8799cd's 3--5 $\mu m$ measurements predict far too flat an SED from $L^\prime$ to [4.05].  The \citet{Marley2012} models were also compared to HR 8799 bcd photometry and spectra to constrain the planets' masses, radii, and cloud properties.  However,  comparing the planets' SEDs (e.g. Figure \ref{seds2}) to their best-fit models (from their Figures 4, 5, and 6) shows that the models do not match HR 8799 bcd's photometry at multiple mid-IR passbands.    

Models invoking thicker (i.e. A60 instead of AE60), perhaps ``patchy" clouds may provide better matches to HR 8799 bcde photometry and, in some cases, modify our interpretations of the planets' SEDs.   Based on the planets' 3.3 $\mu m$ and $M^\prime$ photometry, several studies argue that the planets exhibit evidence for non-equilibrium carbon chemistry \citep{Hinz2010,Galicher2011,Skemer2012,Skemer2014}.   However, only HR 8799 b's SED cannot be matched by any chemical equilibrium models we have considered.  While HR 8799 c's $K$-band spectrum shows clear evidence for abundant $CO$ and weak $CH_{4}$ \citep{Konopacky2013}, HR 8799 de currently lack published $K$-band spectra capable of revealing the same signs of non-equilibrium carbon chemistry.  

Finally, our analyses investigate whether the HR 8799 planets really have different photometric properties.  Using a nominal set of photometry, HR 8799 b and c are clearly different.  In contrast, although HR 8799 c may be slightly fainter than HR 8799 de, the colors for all three planets are as-yet indistinguishable.   We obtain the same general trends using a different source for HR 8799 bcd's $H$-band photometry; using P1640's $J$ and $H$ photometry we can detect no color differences between any of the planets at the 95\% confidence limit or higher.

The strong similarity between HR 8799 cde based on photometry stands in contrast to the P1640 results from \citet{Oppenheimer2013}.  From P1640 $H$-band spectra, they find statistically significantly differences in HR 8799 cde's spectral shapes.  The planets' dissimilar spectra may be due to chemical heterogeneities.  For example, \citet{Oppenheimer2013} report a detection of methane in the spectra of HR 8799 de, but not HR 8799 bc.   However, these results are challenging to understand if the planets' thermal IR 3.3 $\mu m$ and $M^\prime$ photometry are due to non-equilibrium carbon chemistry favoring methane depletion and enhanced $CO$  \citep{Skemer2012,Skemer2014}.    On the other hand, the identifications/non-detections of molecular features in the P1640 spectra are ``tentative";  chemical diversity in the HR 8799 planets they find may be a natural consequence of the diversity in chemical environment of the protoplanetary disk surrounding HR 8799 when the planets formed \citep[e.g.][]{Oberg2011}.

New, high SNR spectra of HR 8799 cde at multiple near-IR wavelengths are required to better assess whether HR 8799 cde are chemically heterogeneous despite having nearly identical photometric properties, luminosities and masses.   In addition to new, higher Strehl/contrast P1640 data, integral field spectroscopy with the \textit{Gemini Planet Imager} \citep{Macintosh2014}, SPHERE \citep{Beuzit2008} or SCExAO/CHARIS \citep{Martinache2009,Peters2012} should yield high enough SNR detections of the planets to decisively tell whether or not the planets have different-shaped spectra indicative of chemical heterogeneities.

\acknowledgements 
The \textit{NASA-Keck}, \textit{ESO} and \textit{Subaru} Time Allocation Committees supported this work through generous allotments of observing time.
We thank Christian Marois and Laurent Pueyo for extensive discussions focused on advanced high-contrast image processing techniques, 
Stanimir Metchev for discussions on NIRC2 linearity corrections, Adam Burgasser for suggestions on comparing our data to field substellar objects, 
Ernst De Mooij for helpful discussions on flux calibration strategies, Timothy Brandt for help regarding recent research results, M. Xu for moral support, and the anonymous referee for helpful suggestions that improved the quality of this paper.
Some of the data presented herein were obtained at the W.M. Keck Observatory, which is operated as a scientific partnership among the California Institute of Technology, the University of California and the National Aeronautics and Space Administration. The Observatory was made possible by the generous financial support of the W.M. Keck Foundation. 
This work is based on data collected at Subaru Telescope, which is operated by the National Astronomical Observatory of Japan.
The authors wish to recognize and acknowledge the very significant cultural role and reverence that the summit of Mauna Kea has always had within the indigenous Hawaiian community.  We are most fortunate to have the opportunity to conduct observations from this mountain. 
This research has made use of the Keck Observatory Archive (KOA), which is operated by the 
W. M. Keck Observatory and the NASA Exoplanet Science Institute (NExScI), under contract with the National Aeronautics and Space Administration. 
We are extremely grateful to the NExScI/KOA staff for developing and maintaining
 the NIRC2 archive.  TC is partially supported by a McLean Postdoctoral Fellowship.

{}
\begin{deluxetable}{lcllccccccc}
\setlength{\tabcolsep}{0pt}
\tablecolumns{8}
\tablecaption{Observing Log}
\tiny
\tablehead{{UT Date}&{Telescope/Instrument}&{Pixel Scale}&{Filter}&{$t_{int}$}&{$N_{images}$}&{$\Delta$PA} \\
{} & {} &{(mas pixel$^{-1}$)} &{}&{(s)}&{}&{(degrees)}}
\startdata
\textit{New Data}\\
2012-07-29 & {Subaru/IRCS} & 20.43&$L^\prime$ & 30 & 97 & 173.4 \\
2012-08-25 & {VLT/NaCo} & 27.1&{$L^\prime$}&{40}&{188}&{61.5} \\
2012-08-26 & {VLT/NaCo} & 27.1&{[4.05]}&{40}&{238}&{56.6} \\
2012-10-30 & {Keck/NIRC2} & 9.952&{$L^\prime$}&{21}&{171}&{171.1} \\
\textit{Archival Data}\\
2010-07-21 & {Keck/NIRC2} & 9.952 & {$L^\prime$} & {25} & {63} & {155.2} \\
 \enddata
\tablecomments{The archival 2010 Keck data were published in \citet{Marois2010a}.}
\label{hr8799log}
\end{deluxetable}

\begin{deluxetable}{lcllccccccc}
\setlength{\tabcolsep}{0pt}
\tablecolumns{8}
\tablecaption{Detections and Photometry}
\tiny
\tablehead{{Planet}&{Filter}&{Telescope/Instrument}&{SNR}&{$\Delta$m}&{Abs. Mag}}
\startdata
\hline
\\
HR 8799 b & $L^\prime$ & Subaru/IRCS & 8.7 & 10.43 $\pm$ 0.16 & 12.64 $\pm$ 0.16\\
          &           & Keck/NIRC2-2012 & 43 & 10.26 $\pm$ 0.10 & 12.50 $\pm$ 0.10\\
          &           & Keck/NIRC2-2010 & 22 & 10.36 $\pm$ 0.10 & 12.60 $\pm$ 0.10\\
          &           & VLT/NaCo & 31 & 10.30 $\pm$ 0.11 & 12.54 $\pm$ 0.11\\
          & [4.05]    & VLT/NaCo & 7 & 9.60 $\pm$ 0.18 & 11.84 $\pm$ 0.18\\
          \hline
\\
HR 8799 c & $L^\prime$ & Subaru/IRCS & 12 & 9.47 $\pm$ 0.14 & 11.71 $\pm$ 0.10\\
          &           & Keck/NIRC2-2012 & 36 & 9.44 $\pm$ 0.08 & 11.68 $\pm$ 0.08\\
          &           & Keck/NIRC2-2010 & 34 & 9.50 $\pm$ 0.08 & 11.74 $\pm$ 0.08\\
          &           & VLT/NaCo & 37 & 9.43 $\pm$ 0.11 & 11.67 $\pm$ 0.11 \\
          & [4.05]    & VLT/NaCo & 13 & 8.75 $\pm$ 0.11 & 10.99 $\pm$ 0.08\\
          \hline
\\
HR 8799 d & $L^\prime$ & Subaru/IRCS & 10 & 9.30 $\pm$ 0.14 & 11.54 $\pm$ 0.12\\
          &           & Keck/NIRC2-2012 & 26 & 9.29 $\pm$ 0.09&11.53 $\pm$ 0.09\\
          &           & Keck/NIRC2-2010 & 19 &  9.34 $\pm$ 0.09 & 11.58 $\pm$ 0.09\\
          &           & VLT/NaCo & 14 & 9.33 $\pm$ 0.14 & 11.57 $\pm$ 0.14\\
          & [4.05]    & VLT/NaCo & 8.7 & 8.65 $\pm$ 0.15 & 10.89 $\pm$ 0.13\\
          \hline
\\
HR 8799 e & $L^\prime$ & Subaru/IRCS & 5.5 & 9.23 $\pm$ 0.24 & 11.47 $\pm$ 0.25\\
          &           & Keck/NIRC2-2012 & 15 & 9.27 $\pm$ 0.15& 11.51 $\pm$ 0.15\\
          &           & Keck/NIRC2-2010 & 10 & 9.33 $\pm$ 0.12 & 11.57 $\pm$ 0.12\\
          &           & VLT/NaCo & 8.1 &9.27 $\pm$ 0.21 & 11.51 $\pm$ 0.21\\
          & [4.05]    & VLT/NaCo & 6.3 & 8.50 $\pm$ 0.20 & 10.74 $\pm$ 0.20\\
 \enddata
\tablecomments{The two Keck $L^\prime$ entries for each planet refer to new data (2012) and archival data (2010), respectively.}
\label{hr8799phot}
\end{deluxetable}

\begin{deluxetable}{lcllccccccc}
\setlength{\tabcolsep}{0pt}
\tablecolumns{8}
\tablecaption{Astrometry}
\tiny
\tablehead{{Planet}&{2010-07-21 ([E,N]\arcsec{})}&{2012-10-30 ([E,N]\arcsec{})}}
\startdata
b &  [ 1.547 $\pm$ 0.006,  0.757 $\pm$ 0.009] & [1.558 $\pm$ 0.006, 0.729 $\pm$ 0.009]\\
c &  [-0.606 $\pm$ 0.006,  0.725 $\pm$ 0.006] & [-0.557 $\pm$ 0.006, 0.763 $\pm$ 0.006]\\
d &  [-0.269 $\pm$ 0.006, -0.580 $\pm$ 0.006] & [-0.343 $\pm$ 0.006, -0.555 $\pm$ 0.006]\\
e &  [-0.329 $\pm$ 0.006, -0.178 $\pm$ 0.006] & [-0.371 $\pm$ 0.006, -0.080 $\pm$ 0.006]
 \enddata
\tablecomments{The astrometric errors assume a centroiding accuracy of 0.5 pixels ($\sim$ 0\farcs{}005 for Keck/NIRC2) and consider the intrinsic SNR of the detection and the uncertainty in the north position angle.}
\label{hr8799astrom}
\end{deluxetable}
\begin{deluxetable}{llllllllllllll}
 \tiny
\setlength{\tabcolsep}{0.0001in}
\tabletypesize{\scriptsize}
\tablecolumns{10}
\tablecaption{Nominal HR 8799 \lowercase{bcde} Photometry Considered Here}
\tablehead{{Filter}&{$z$}&{$J$}&{$H$}&{$K_{s}$}&{[3.3]}&{$L^\prime$}&{[4.05]} & {$M^\prime$}\\
{$\lambda$ ($\mu m$)}&{1.03} & {1.25 } & {1.633}&{2.146}&{3.3}&{3.776}&{4.051} & {4.68}}
\startdata
Planet\\
b&18.24 $\pm$ 0.29 & 16.52 $\pm$ 0.14 & 15.08 $\pm$ 0.13 &  14.05 $\pm$ 0.08 & 13.22 $\pm$ 0.11&12.60 $\pm$ 0.10 & 11.84 $\pm$ 0.18 & 13.07 $\pm$ 0.30\\
c&$>$ 16.48 & 14.65 $\pm$ 0.17 & 14.18 $\pm$ 0.17 & 13.13 $\pm$ 0.08 & 12.22 $\pm$ 0.11& 11.74 $\pm$ 0.08 & 10.99 $\pm$ 0.08 & 12.05 $\pm$ 0.14\\
d&$>$15.03 & 15.26 $\pm$ 0.43 & 14.23 $\pm$ 0.22 &  13.11 $\pm$ 0.12 & 12.02 $\pm$ 0.11&11.58 $\pm$ 0.09 & 10.89 $\pm$ 0.14 & 11.67 $\pm$ 0.35\\
e& -- & $>$ 13.2 & 13.88 $\pm$ 0.22 &12.89 $\pm$ 0.26& 12.02 $\pm$ 0.21 & 11.57 $\pm$ 0.12 & 10.74 $\pm$ 0.20 & $>$ 10.09\\
\enddata
\tablecomments{The (nominal) HR 8799 bcde photometry considered here.  The $z$ band photometry come from \citet{Currie2011a}, 
$J$ band photometry for HR 8799 bcd come from \citet{Marois2008} while upper limits come from \citet{Oppenheimer2013}, the $H$ band and [3.3] photometry come from 
\citet{Skemer2012}, the $K_{s}$ band photometry come from \citet{Marois2008} for HR 8799 bcd and from \citet{Currie2011a} for HR 8799 e, the 
$L^\prime$ and [4.05] photometry come from this work and the $M^\prime$ photometry come from \citet{Galicher2011}.
} 
\label{tab_litphot}
\end{deluxetable}

\begin{deluxetable}{llllllllllllll}
 \tiny
\setlength{\tabcolsep}{0.0001in}
\tabletypesize{\tiny}
\tablecolumns{10}
\tablecaption{Statistical Tests Comparing HR 8799 bcde photometry}
\tablehead{{Photometry} & {b and c} & {b and d } & {b and e} & {c and d} & {c and e} & {d and e}}
Nominal & (0.98 [0.42], 0.99+) & (0.32 [0.39], 0.99+ ) & (0.14 [0.36], 0.99+) & (0.59 [0.92],0.66) & (0.02 [0.82], 0.66) & (0.27 [0.91], 0.31)\\
Keck 2005 H band & (0.98 [0.42], 0.99+) & (0.42 [0.38], 0.99+) & (0.14 [0.36], 0.99+) & (0.64 [0.90],0.77) & (0.01 [0.83], 0.59) & (0.19 [0.96], 0.11)\\
P1640 J and H band & (0.10 [0.43], 0.99+) & (0.58 [0.38], 0.99+) & (0.74 [0.35], 0.99+) & (0.35 [0.87],0.77) & (0.64 [0.80], 0.92) & (0.27 [0.92], 0.28)\\
\enddata
\tablecomments{
Confidence limits at which two pairs of planets' (scaled) photometry differ 
as determined from the goodness-of-fit statistic given the number of degrees of freedom -- 
$N_{data}$ - 2 for scaled photometry, $N_{data}$ -1 for non-scaled photometry -- and 
the $\chi^{2}$ value.  For HR 8799 bcd comparisons, $N_{data}$ = 7 while comparisons with HR 8799 e have $N_{data}$ = 5.  
The different rows indicate the nominal case, using photometry listed in Table \ref{tab_litphot}, the \citet{Currie2012b} $H$-band photometry, 
and the \citet{Oppenheimer2013} $J$ and $H$ photometry.  The two entries (enclosed by parentheses) list the correlation from scaling the planets' fluxes with respect to one another (left entry) and the nominal correlation (no scaling of the spectra) (right entry).  
A value closer to one means that the photometry significantly differ (e.g. 0.95 = the planets' photometry differ at the 2-$\sigma$/95\% confidence limit).  
The ``[]" values enclose the scaling factor applied to the second planet in each listed pair that minimizes $\chi^{2}$.}
\label{tab_var}
\end{deluxetable}

\begin{deluxetable}{ccccccccccccc}
\tabletypesize{\scriptsize}
\tablecaption{Atmospheric Models}
\tablewidth{0pt}
\tablehead{
\colhead{Figure} &
\colhead{Panel} &
\colhead{Planet} &
\colhead{T$_{eff}$(K)} &
\colhead{log(g)} &
\colhead{Cloud Type} &
\colhead{Chemistry} &
\colhead{Reference}
}

\startdata
\ref{seds1}    & top-left & HR 8799 b & 900  & 4.0 & A60 & equilibrium & 1,2   \\
                        & middle-left  &"&    850  &  4.0   &  AE30  & equilibrium  & 3  \\
                         & bottom-left &"&  850   &   4.3  & AE60  & 100$\times$CO, 0.01$\times$CH$_{4}$&4  \\
                          & top-right  &HR 8799 d&    1000  &    4.0 &   A60    & equilibrium &1,2  \\
                         & middle-right  &"&   975   &    4.0 &  AE60     & equilibrium  &3 \\
                         & bottom-right&"&  1000 &  4.0   &  AE60     & 100$\times$CO, 0.01$\times$CH$_{4}$ &4 \\

\hline
\ref{seds2}  & top-left& HR 8799 b & 900 & 4.0 & 0.85$\times$A60, 0.15$\times$E60 & equilibrium  & 5\\                         
                          & bottom-left &HR 8799 c&    1000  &  4.0   &  0.7$\times$A60, 0.3$\times$E60  & equilibrium  & 5  \\
                          & top-right &HR 8799 d&    1000  &    4.0 &   0.9$\times$A60, 0.1$\times$AE60    & equilibrium &5 \\
                      & bottom-right&HR 8799 e&  1000 &  4.0   &  0.75$\times$A60, 0.25$\times$E60    & equilibrium &5 \\
\enddata
\tablecomments{References: 1) \citet{Burrows2006}, 2) \citet{Currie2011a}, 3) \citet{Madhusudhan2011}, 4) \citet{Skemer2012}, 
5) this work.
}
\label{modelused}
\end{deluxetable}


\end{document}